\documentclass[aps,prd,showpacs,notitlepage,nofootinbib,superscriptaddress,floatfix,showkeys,twocolumn]{revtex4-2}

\usepackage[T1]{fontenc} 
\usepackage[utf8]{inputenc}
\usepackage{pstool}
\usepackage{hyperref}
\usepackage{amsmath,amssymb}
\usepackage{float}
\usepackage{graphicx}
\usepackage{graphics}
\usepackage{bm}
\usepackage{latexsym}
\usepackage{psfrag}
\usepackage{color}
\usepackage{subfigure}
\usepackage[normalem]{ulem}
\usepackage{textcomp}
\usepackage{comment}
\usepackage{braket}
\usepackage{mathalpha}
\usepackage{tabularx}
\usepackage{placeins}   
\usepackage{xcolor}
\usepackage{physics}
\usepackage{simpler-wick}
\usepackage{mathtools}
\usepackage{ulem}

\def\bea{\begin{eqnarray}}
\def\eea{\end{eqnarray}}
\def\nn{\nonumber}

\def\f{\frac}

\def\d{{\rm d}}

\def\mpcinv{{\rm Mpc}^{-1}}

\def\vk{{\bm{k}}}
\def\vq{{\bm{q}}}
\def\hvk{{\hat{\bm{k}}}}

\def\cR{{\cal R}}

\def\ps{\mathcal{P}_{\mathcal{R}}}

\def\ph{\mathcal{P}_{h}}
\def\barpow1d{\mathcal{P}^{1\rm D}_{\mathrm{b}}}

\def\ns{n_{_{\mathrm{S}}}}
\def\As{A_{_{\rm S}}}

\def\ogw{\Omega_{_{\mathrm{GW}}}}
\def\fnl{f_{{\rm NL}}}
\def\taunl{\tau_{   {\rm NL}}}

\allowdisplaybreaks

\begin{document}
\title{Constraining primordial non-Gaussianity and parity-violation through Scalar-Induced Gravitational Waves with next-generation ground-based interferometers}
\author{Ilaria~Caporali}
\email{ilaria.caporali@phd.unipi.it}
\affiliation{Dipartimento di Fisica ``Enrico Fermi'', Università di Pisa, Largo Bruno Pontecorvo 3, Pisa I-56127, Italy}
\affiliation{Istituto Nazionale di Fisica Nucleare (INFN), Sezione di Pisa, Largo Bruno Pontecorvo 3, Pisa I-56127, Italy}
\author{H.~V.~Ragavendra}
\email{ragavendra.hv@pd.infn.it}
\affiliation{Dipartimento di Fisica e Astronomia “Galileo Galilei”, Universit\`{a} degli Studi di Padova, Via Marzolo 8, I-35131 Padova, Italy}
\affiliation{Istituto Nazionale di Fisica Nucleare (INFN), Sezione di Padova, Via Marzolo 8, I-35131 Padova, Italy}
\author{Angelo Ricciardone}
\email{angelo.ricciardone@unipi.it}
\affiliation{Dipartimento di Fisica ``Enrico Fermi'', Università di Pisa, Largo Bruno Pontecorvo 3, Pisa I-56127, Italy}
\affiliation{Istituto Nazionale di Fisica Nucleare (INFN), Sezione di Pisa, Largo Bruno Pontecorvo 3, Pisa I-56127, Italy}
\author{Nicola Bartolo}
\email{nicola.bartolo@pd.infn.it}
\affiliation{Dipartimento di Fisica e Astronomia “Galileo Galilei”, Universit\`{a} degli Studi di Padova, Via Marzolo 8, I-35131 Padova, Italy}
\affiliation{Istituto Nazionale di Fisica Nucleare (INFN), Sezione di Padova, Via Marzolo 8, I-35131 Padova, Italy}
\affiliation{Istituto Nazionale di Astrofisica (INAF), Osservatorio Astronomico di Padova, vicolo dell’Osservatorio 5, I-35122 Padova, Italy}
\begin{abstract}
In this work, we investigate the prospects for probing primordial non-Gaussianity and 
associated symmetry of parity through scalar-induced gravitational waves (SIGWs), 
with third-generation gravitational-wave detectors. 
We develop a framework that accounts for contributions to the energy density spectrum of GWs arising from the 
scalar non-Gaussianity quantified by bispectrum and trispectrum,
and perform parameter inference using simulated data from Einstein Telescope and Cosmic Explorer. 
The parity-odd component of the scalar trispectrum induces circular polarization in the stochastic gravitational-wave background (SGWB), providing a direct probe of parity-violation in the primordial Universe. 
We show that future interferometers can place competitive constraints on the parity-odd scalar trispectrum, along 
with the bispectrum and parity-even trispectrum. 
Moreover, we include the astrophysical contribution, which could act as a foreground for the SIGWs.
We show that, despite the addition of such a foreground, we are still able to effectively constrain 
the cosmological parameters related to SIGWs and the astrophysical parameters as well.
\end{abstract}
\maketitle

\section{Introduction}

An irreducible cosmological source of gravitational waves arise from scalar (i.e., density) perturbations feeding 
the tensor metric fluctuations 
at non-linear order in perturbation theory~\cite{Tomita:1967wkp,Matarrese:1992rp,Matarrese:1993zf,Matarrese:1996pp,
Matarrese:1997ay,Ananda:2006af,Baumann:2007zm,Bartolo:2007vp,Guzzetti:2016mkm,Kohri:2018awv,Domenech:2021ztg}.
This phenomenon of scalar-induced gravitational waves (SIGWs) 
is a prediction of General Relativity and, a priori, is not relying on any specific model.
Since we have precise constraints on the power spectrum of cosmological scalar perturbations over large scales from cosmic microwave background (CMB) anisotropies, SIGWs represent an irrefutable prediction of precision cosmology.
SIGWs became very popular given their connection with the production of primordial black holes (PBHs), which could constitute a significant fraction of dark matter today (see, for 
instance~\cite{Carr:2009jm,Carr:2016drx,Bartolo:2016ami,Carr:2017jsz,Bartolo:2018evs,Bartolo:2018rku,Carr:2020gox,Carr:2020xqk,LISACosmologyWorkingGroup:2023njw,Ozsoy:2023ryl,Ragavendra:2023ret} and references therein).
They have been proposed also as potential candidates to explain the GW signals detected by the Pulsar Timing Collaboration 
(PTA) in the nano-Hertz regime~\cite{NANOGrav:2023hvm, Cai:2023dls, Franciolini:2023pbf, Chang:2023vjk, Chang:2023aba, Li:2023xtl,
Firouzjahi:2023lzg, Yi:2023tdk, Liu:2023ymk, Domenech:2024rks, Wang:2025kbj,
Gouttenoire:2025jxe}. Forecasts about their detectability have been performed also at higher frequencies, from the milli-Hertz, where the future Laser Interferometer Space Antenna (LISA) will operate 
~\cite{LISACosmologyWorkingGroup:2025vdz}, to tens of Hertz, through LIGO-Virgo-KAGRA networks ~\cite{Kapadia:2020pnr,Inui:2023qsd, Martinovic:2021hzy, LIGOScientific:2025kry} and future third-generation ones, as Einstein Telescope (ET)~\cite{Punturo:2010zz,ET:2019dnz} and Cosmic Explorer (CE)\cite{Reitze:2019iox}.

The presence of chirality, i.e., circular polarization, in SIGWs, has been recently addressed in \cite{Ragavendra:2025svk}, as a tell-tale signature of parity-violation in the primordial universe.
That study shows that a parity-odd primordial scalar trispectrum, even with a modest amplitude, can induce a significant degree of chirality in SIGWs.
The mechanism establishes complementarity between the predictions of the parity-odd scalar trispectrum and the associated chirality of SIGW arising from an underlying parity-violating model of early universe.
Several efforts in the literature have analysed and constrained the scalar non-Gaussianity using the ansatz of local template, through their respective contributions to SIGWs~\cite{Cai:2018dig,Unal:2018yaa,Adshead:2021hnm,Garcia-Saenz:2022tzu,Liu:2023ymk,Li:2023xtl,Chang:2023vjk,
Chang:2023aba,Yuan:2023ofl,Perna:2024ehx,Li:2025met}.
A Fisher analysis performed in~\cite{Perna:2024ehx}, together with a Bayesian analysis in~\cite{LISACosmologyWorkingGroup:2025vdz}, demonstrated the ability of SIGWs to levy efficient constraints on primordial non-Gaussianity parameters associated with the local template in the LISA frequency band.

In this work, we generalize the approach by relaxing the ansatz of local shape for primordial
non-Gaussianity while also accounting parity-odd correlations at the level of trispectrum.
We work with the signal of SIGWs arising from scalar power that is peaked over 
scales probed by next generation GW interferometers, namely ET in its 2L aligned configuration with $15 \, \rm{km}$ long arms \cite{Branchesi:2023mws, ET:2025xjr} and 
CE \cite{Evans:2023euw} (placed in the location of LIGO Hanford, considering a configuration with $40 \, \rm km$ long arms). 
We compare the spectra $\ogw^I$ and $\ogw^V$ against simulated data of ET and CE, and 
obtain constraints on the primordial non-Gaussianity parameters, including the strength of parity-odd trispectrum.

To add context to this work with respect to other observational efforts surrounding cosmological parity-violation, we may note the growing evidence of cosmic birefrigence in CMB~\cite{Minami:2020odp,
Diego-Palazuelos:2022dsq, Komatsu:2022nvu, Greco:2022xwj, Arcari:2024nhw, Gruppuso:2025ywx,Diego-Palazuelos:2025dmh} and the emerging constraints on the parity-odd scalar trispectrum from galaxy distribution~\cite{Hou:2022wfj, Philcox:2022hkh, Krolewski:2024paz, Hewson:2024rnb,Kurita:2025hmp}.
These are parity-violating signatures over large scales of several megaparsecs.
Our analysis probes cosmological parity-violation over extremely small scales of around $10^6$ meters.
The constraint on parity-odd primordial scalar trispectrum over small scales shall complement the constraint from large scales and inform us about its intrinsic scale dependence.
In terms of inflationary dynamics, this analysis is a window to test parity in the scalar sector over the last few e-folds of evolution before reheating begins.

The paper is organized as follows. In Section \ref{sec:The Stochastic Gravitational Wave Background} we briefly introduce the Stochastic Gravitational Wave Background (SGWB) and the relation with the two-point correlation function, also for chiral GWs. In Section \ref{sec:Scalar-induced gravitational waves to probe primordial parity-violation} we  
discuss the mechanism of SIGWs, accounting for primordial scalar non-Gaussianity and associated parity-violation. 
In Section \ref{sec:Bayesian Analysis for the SGWB} we present the Bayesian framework adopted for the analysis, 
detailing the data generation, the likelihood, and discussing the prior choices. We finally present our results in Section \ref{sec:Results}, showing that we are not only able to constrain the primordial non-Gaussianity parameters, 
but also the chirality of the phenomenon. 
We also illustrate the results in terms of the non-trivial shapes of the degree of polarization. 
Further in Section \ref{sec:Discussion}, we discuss the theoretical implications of our results 
and conclude in Section \ref{sec:Conclusion} by summarizing the significance, novelty and outlook of our findings.

\section{The Stochastic Gravitational Wave Background}
\label{sec:The Stochastic Gravitational Wave Background}

The SGWB can be described as a superposition of plane waves, expanded as 
\begin{equation}
    h_{ij}(t, \vec{x}) = \int df \, d^2{\Omega}_{\hat{k}} \sum_{p} \tilde{h}_{p}(f, \hat{k}) \, e_{ij}^{p}(\hat{k}) \, e^{i 2 \pi f (t - \hat{k}\cdot \vec{x}/c)}\,.
\label{eq:hij}
\end{equation}
Here, $\hat{k}$ is the propagation direction, $p= +, \times$ labels the two tensor polarizations, $e_{ij}^{p}$ are the corresponding polarization tensors, and $\tilde{h}_{p}(f,\hat{k})$ denotes the Fourier amplitude of the background. 

For a SGWB that is stationary, isotropic, and unpolarized, the two-point correlation function can be written as
\begin{equation}
    \langle \tilde{h}_p(f, \hat{k}) \tilde{h}_{p'}^*(f', \hat{k}')\rangle = \frac{\delta(f-f')}{2} \frac{\delta_{\hat{k} \hat{k}'}}{4\pi} \delta_{pp'} \frac{3 H_0^2}{4 \pi^2 f^3} \Omega_{\rm GW}(f) \, ,
    \label{eq:twopoint_corr}
\end{equation}
where $\delta_{pp'}$ means that the two polarizations are uncorrelated, $\delta_{\hat{k} \hat{k}'}$ the isotropy, and $\delta(f-f')$ the stationarity of the background. $H_0$ is the Hubble parameter today $H_0 = 67.4 \rm\,  km \, s^{-1} \, Mpc^{-1}$ \cite{Planck:2018vyg} and $\Omega_{\rm GW}$ represents the GW energy density per logarithmic interval of frequency.

The projected GW amplitude on the detector $i$, $\tilde{h}_i(f)$, is related to the GW amplitude $\tilde{h}_p$ through the detector's pattern function $F_i^\lambda$
\begin{equation}
    \tilde{h}_i(f) = \int d^2{\Omega}_{\hat{k}} \sum_p F_i^{p}(f,\hat{k})\, \tilde{h}_p(f,\hat{k}) \, .
    \label{eq:GWsignal_Fourier}
\end{equation}

The SGWB signal has a zero mean and its covariance given by~\cite{Allen:1997ad, Romano:2016dpx}
\begin{equation} 
\langle \tilde{h}_i(f)\tilde{h}^*_j(f^\prime)\rangle \equiv \frac{\delta(f-f^\prime)}{2}\frac{3H_0^2}{10\pi^2f^3} \, \gamma_{ij}^t(f) \,  \Omega^I_{\rm GW}(f) \, , \label{eq:2pt} 
\end{equation}
where the quantity $\gamma_{ij}^t(f)$ represents the normalized overlap reduction function (ORF) for tensor modes, defined as \cite{Thrane:2013oya, Romano:2016dpx}
\begin{equation}
\gamma_{ij}^t(f) \equiv \frac{5}{8\pi} \int d^2{\Omega}_{\hat{k}} \sum_p F_i^p(f, \hat{k}) F_j^{p, *}(f, \hat{k})  e^{-2\pi i f \hat{k} \cdot (\vec{x}_i - \vec{x}_j)/c} \, . 
\label{eq:orf_general_basis}
\end{equation} 
The indices $i,j$ correspond to the interferometers considered. 
In the case of tensor modes, the ORF is normalized to unity for two co-located, L-shaped detectors.
The notion of ORF can also be extended to chiral SGWB as \cite{Romano:2016dpx, Caporali:2025dyf}

\begin{align}
\gamma_{ij}^V(f) \equiv& -\frac{5i}{8\pi} \int d^2{\Omega}_{\hat{k}}  \left( F_i^+(f, \hat{k}) F_j^{\times, *}(f, \hat{k}) \right. \nonumber  \\
& \left . -  F_i^{\times}(f, \hat{k}) F_j^{+, *}(f, \hat{k})\right) e^{-2\pi i f \hat{k} \cdot (\vec{x}_i - \vec{x}_j)/c} \, . 
\label{eq:orf_general_basis_V}
\end{align}

\subsection{Two-point correlation function for chiral GW}
\label{subsec: chiral GW}

For signals containing a parity-violating component, the two-point correlation function of the projected GW amplitudes is
\begin{align}
    \langle\tilde{h}_i(f)\tilde{h}^*_j(f^\prime)\rangle &\equiv \frac{\delta(f-f^\prime)}{2}\frac{3H_0^2}{10\pi^2f^3} \, \nonumber \\ 
    &\left( \gamma_{ij}^t(f) \,  \Omega^I_{\rm GW} (f) + \gamma_{ij}^V(f) \,  \Omega^V_{\rm GW} (f) \right) \, , \label{eq:2pt_IV} 
\end{align}
where $\gamma_{ij}^t(f)$ is the ORF for tensor modes, defined in Eq.\eqref{eq:orf_general_basis}, while $\gamma_{ij}^V(f)$ is the ORF for the V Stokes parameter (see \cite{Caporali:2025dyf} for a detailed discussion), $\Omega^I_{\rm GW} (f)$ is the intensity contribution to the energy density spectrum, as defined in Eq.\eqref{eq:2pt} and $\Omega^V_{\rm GW} (f)$ is the parity violating part of the spectrum, 
that quantifies the difference between left and right chiral components of the spectrum.

\subsection{Characterization of the noise}
\label{subsec:noise}

Considering two detectors $i$ and $j$,
the covariance of the noise is defined as the two-point correlation function
\begin{equation}
    \left\langle \tilde{n}_i(f)\tilde{n}^*_j(f^\prime)\right\rangle \equiv \frac{\delta(f-f^\prime)}{2} N_{ij}(f) \, ,  
    \label{def:PSD_noise}
\end{equation}
where we have assumed that the noise is stationary and, therefore, not correlated at different frequencies.
In Eq.~\eqref{def:PSD_noise}, the elements along the diagonal, i.e., those with $i=j$, define the Power Spectral Density (PSD) and represent the auto-correlation functions of the detector noise. The elements outside the diagonal, i.e., those with $i \neq j$, define the Cross Power Spectral Density (CSD) and quantify a possible presence of correlated noise between different detectors. In this work, we neglect any possible source of correlated noise among different detectors. See \cite{Caporali:2025mum} for an analysis of the SGWB detection prospects in presence of correlated noise.

\section{Scalar-induced gravitational waves to probe primordial parity-violation}
\label{sec:Scalar-induced gravitational waves to probe primordial parity-violation}

Having quantified the two-point correlation of the GW signal as seen by the detectors,
we shall turn to model the signal of SIGWs.
A remarkable property of SIGWs is its sensitivity to scalar non-Gaussianity~\cite{Unal:2018yaa,Cai:2018dig,Adshead:2021hnm,Garcia-Saenz:2022tzu,Perna:2024ehx}.
When the scalar perturbations are enhanced to large amplitudes, the associated non-Gaussianity, quantified by correlation functions beyond the power spectrum, is proportionately enhanced. 
The non-Gaussian components of the scalar perturbations, quantified by the bispectrum and trispectrum, 
induce additional contributions to the SIGW.
In fact, the analyses of SIGWs from concrete inflationary models reveals that such scalar non-Gaussian contributions affect the spectral energy density of SIGWs~\cite{Ragavendra:2020vud,Ragavendra:2021qdu,Garcia-Saenz:2023zue}. 

Several models of inflation that involve parity-violating interactions in their description are known to produce parity-odd component in the scalar trispectrum~\cite{Cabass:2022rhr,Niu:2022fki,Fujita:2023inz,Reinhard:2024evr,Moretti:2024fzb,Orlando:2025fec,Yura:2025mus,Cho:2025rvg}. 
A parity-odd trispectrum shall have asymmetric amplitude between its configurations that are related by the inversion of coordinates.
When accounting for the complete form of scalar trispectrum, the parity-odd part of it, if present, shall induce a difference between the right and left chiral modes of SIGW.
This results in the V-mode of SIGW $\ogw^V$ that is directly proportional to the strength of the
parity-odd component~\cite{Ragavendra:2025svk}.

We should also mention that parity-violating models may generate chiral features in the primary tensor perturbations, wherein the primary GW acquires a V-mode to its two-point correlation, see, e.g.~\cite{
Sorbo:2011rz,Anber:2012du,Adshead:2013qp,Caprini:2014mja,Dimastrogiovanni:2016fuu,Dimastrogiovanni:2025snj}.
However, it is a model-dependent effect. 
For instance, models with gravitational Chern-Simons coupling predict the chirality of primary GW 
to be highly
suppressed~\cite{Satoh:2010ep,Bartolo:2017szm,Bartolo:2018elp,Qiao:2019hkz,Bartolo:2020gsh,Creque-Sarbinowski:2023wmb,Moretti:2024fzb,Orlando:2025fec}. 
On the other hand, models involving axion fields coupled to gauge fields predict their primary GW 
to be maximally chiral 
(for instance~\cite{Sorbo:2011rz,Adshead:2013qp,Caprini:2014mja,Dimastrogiovanni:2016fuu,
Dimastrogiovanni:2025snj}).

In a scenario where SIGWs are enhanced by amplification of scalar power, the primary tensor perturbations become sub-dominant with respect to the scalar-induced tensor perturbations 
(see for examples~\cite{Braglia:2020eai,Bhaumik:2019tvl,Ragavendra:2020sop,Ragavendra:2023ret}).
Hence, the contributions of primary GW to the total $\ogw^I$ and $\ogw^V$ shall be negligible 
against those of SIGWs.

Our theoretical setup closely follows the description presented in~\cite{Ragavendra:2025svk}.
The two-point correlation of the tensor mode-function $\tilde h^\lambda_\vk$ at a given conformal time $\eta$ [as seen in Eq.~\eqref{eq:hij} but in $\eta$ instead of frequency space $f$]
is defined as 
\begin{align}
\langle \tilde h^{\lambda_1}_{\vk_1}(\eta) \tilde h^{\lambda_2}_{\vk_2}(\eta) \rangle
=& (2\pi)^3\f{2\pi^2}{k_1^3} \ph^{\lambda_1\lambda_2}(k,\eta)\,\delta^{(3)}(\vk_1+\vk_2)\,,
\end{align}
where $\ph^{\lambda_1\lambda_2}(k,\eta)$ is the dimensionless power spectrum 
and $\lambda_1$ and $\lambda_2$ are the polarization indices.
We may define the correlation matrix in the basis of circular polarizations 
with $\lambda=[L,R]$ as
\begin{align}
\ph^{\lambda_1\lambda_2} =&
\begin{pmatrix} 
\ph^{LL} & \ph^{LR} \\ 
\ph^{RL} & \ph^{RR}
\end{pmatrix}\,.
\end{align}
If we rewrite the above combinations in terms of spectral densities associated with
the Stokes parameters $I, Q, U$ and $V$, we obtain~\cite{Kato:2015bye}
\begin{align}
\ph^{\lambda_1\lambda_2} =&
\frac{1}{2}\begin{pmatrix} 
\ph^{I}+\ph^{V} & \ph^{Q}-i\ph^{U} \\ 
\ph^{Q}+i\ph^{U} & \ph^{I}-\ph^{V}
\end{pmatrix}\,.
\end{align}
Preserving homogeneity and isotropy leads to vanishing of $\ph^Q$ and
$\ph^U$, i.e. $\ph^{LR}=\ph^{RL}=0$~\cite{Mukherjee:2025dcv}.
This leaves behind the diagonal elements constituted by $\ph^I$ and $\ph^V$.
In the case of violation of parity, the effect of interest that we explore in this
analysis, we shall have the spectral density of $V$, $\ph^V\neq 0$.
Thus the surviving components of the correlation are just
$\ph^I=\ph^{LL}+\ph^{RR}$ and $\ph^V=\ph^{LL}-\ph^{RR}$.

While $\ph^I$ is insensitive to parity, $\ph^V$ captures parity-violation.
Note that although the overall sign of $\ph^V$ is set by the convention in defining the difference, 
any relative sign difference across wavenumbers or time shall be of physical significance.

To relate $\ph^I$ and $\ph^V$ to the sourcing scalar perturbations, let us
recall that $\tilde h^\lambda_\vk$ sourced by the scalar perturbation $\cR$ is 
given by~\cite{Kohri:2018awv,Espinosa:2018eve,Adshead:2021hnm}
\begin{align}
\tilde h^{\lambda}_{\vk}(\eta) =& 4\,\int \f{\d^3 \vq}{(2\pi)^{3}} 
I(q,\vert \vk - \vq \vert, \eta)\, Q^{\lambda}(\vk,\vq) 
\cR_{\vq}\cR_{\vk-\vq}\,,
\label{eq:h-RR}
\end{align}
where $I(q,\vert \vk - \vq \vert, \eta)$ contains the time evolution arising from the Green's function associated with the tensor mode and the transfer
function relating the scalar perturbation during the epoch of sourcing to the primordial
$\cR$~\cite{Adshead:2021hnm,Perna:2024ehx}. 
The factor $Q^\lambda(\vk,\vq)$ is the contraction of the polarization tensor defined 
with respect to the propagating wavevector $\vk$ and the wavevector of the source 
$\vq$~[cf. App.~\ref{app:IQ} for explicit expressions].
Thus the two-point correlation of the scalar-induced tensor perturbations shall be
\begin{align}
\langle \tilde h^{\lambda_1}_{\vk_1}(\eta) \tilde h^{\lambda_2}_{\vk_2}(\eta) \rangle =&
16\iint \f{\d^3 \vq_1 \d^3 \vq_2}{(2\pi)^6} 
I(q_1, \vert \vk_1 - \vq_1 \vert, \eta)\nn \\
& \times\,I(q_2, \vert \vk_2 - \vq_2 \vert, \eta) 
Q^{\lambda_1}(\vk_1,\vq_1) \nn \\
& \times Q^{\lambda_2}(\vk_2,\vq_2)
\langle \cR_{\vq_1}\cR_{\vk_1-\vq_1}\cR_{\vq_2}\cR_{\vk_2-\vq_2}\rangle\,.
\end{align}

In the above expression, the four-point correlation of $\cR$ shall contain the disconnected 
and connected contributions~\cite{Cai:2018dig,Unal:2018yaa,Adshead:2013qp,Ragavendra:2021qdu,Garcia-Saenz:2022tzu,Perna:2024ehx}.
To investigate parity of a given field of perturbations, we look for correlations that transform 
under inversion ($\vk \to -\vk$) distinctly different from transforming under rotation of coordinates.
For scalar perturbations in a statistically isotropic distribution, the transformation of 
two-point and three-point correlations under inversion can always be mapped to rotation and 
hence are insensitive to parity. 
But the transformation of connected four-point correlation, the trispectrum, under inversion is
not always mappable to rotation.
Essentially, the correlations that are not restricted to a plane and three-dimensional are sensitive to 
parity and the lowest order of such a correlation is the trispectrum.

To collect the disconnected and connected contributions to the four-point correlation, we begin with 
the contribution from the power spectra, which is
\begin{align}
\langle \cR_{\vq_1}\cR_{\vk_1-\vq_1}\cR_{\vq_2}\cR_{\vk_2-\vq_2}\rangle=&
2(2\pi)^6\,P_\cR(q_1)P_\cR(\vert\vk_1-\vq_1\vert)\nn \\
& \times \delta^{(3)}(\vq_1+\vq_2) \delta^{(3)}(\vk_1+\vk_2)\,,
\label{eq:4pt-ps2}
\end{align}
where $P_\cR(k)=2\pi^2\ps(k)/k^3$ and $\ps(k)$ is the dimensionless scalar power spectrum.
We have used the symmetry of the integration over $\vq_1$ and $\vq_2$ in writing the above
equation and ignored the contribution that contain $\delta^{(3)}(\vk_1)\delta^{(3)}(\vk_2)$
which does not contribute at any finite wavenumber.

The contribution to the four-point correlation arising from the 
cubic-order action governing scalar perturbations can be computed through the introduction of $\fnl$ in the defining relation of non-Gaussianity in $\cR$ as
$\cR_\vk=\cR_\vk^{\rm G} + \cR^{3\rm nG}_\vk$
and substituting it in the correlation.\footnote{The term $\cR^{3\rm nG}_\vk$ denotes the Fourier mode
of the scalar perturbation arising from a cubic-order interaction in the action governing perturbations.
This interaction leads to the scalar bispectrum whose strength is quantified by $\fnl$ and 
shape is determined by $w_3$~[cf.~App.~\ref{app:3pt-4pt}].}

The expression of $\cR_\vk^{3 \rm nG}$ containing $\fnl$ is given by
\begin{align}
\cR^{3\rm nG}_\vk &= -\f{3}{5}\fnl\int\f{\d^3\vq}{(2\pi)^3}
w_3(k,q,\vert\vk-\vq\vert)\cR_{\vq} \cR_{\vk-\vq}\,,
\end{align}
where the kernel $w_3$ captures the intrinsic scale-dependence of the
scalar bispectrum~\cite{Acquaviva:2002ud,Lyth:2005fi,Schmidt:2010gw,Agullo:2021oqk,Ragavendra:2021qdu}.
This leads to the corresponding four-point correlation being
\begin{align}
\langle \cR_{\vq_1}\cR_{\vk_1-\vq_1}\cR_{\vq_2}\cR_{\vk_2-\vq_2}\rangle =&
8\,(2\pi)^3\,\fnl^2 w_3^2(q_1,q_2,\vert\vq_1-\vq_2\vert) \nn \\
& \times P_\cR(q_2) P_\cR(\vert\vq_1-\vq_2\vert)\nn \\
& \times P_\cR(\vert\vk_1-\vq_1\vert) \delta^{(3)}(\vk_1+\vk_2)\,.
\end{align}

Comparing against Eq.~\eqref{eq:4pt-ps2}, we may note that this term arises from 
loop-level correction arising from the cubic order interaction to the scalar power spectra.
It is neither a fully disconnected term nor a connected term.
Hence it gives rise to the ``hybrid'' contribution to the SIGWs~\cite{Unal:2018yaa,Adshead:2021hnm,Ragavendra:2021qdu,Perna:2024ehx}. 
The other contributions to SIGW that are at the same order of $\fnl^2\ps^3$,
namely the ``C-type'' and ``Z-type'' contributions are in fact contributions that arise 
from the exchange trispectrum~\cite{Adshead:2021hnm,Ragavendra:2021qdu,Perna:2024ehx}. 

Since we model the exchange trispectrum separately, we shall not include those contributions 
here.
We present detailed discussion regarding the bispectrum arising from a cubic-order interaction, 
associated exchange trispectrum, the relation between $\fnl$ and $\taunl$ and the corresponding 
contributions to SIGW in App.~\ref{app:3pt-4pt}.

We then turn to the trispectrum which is the connected part of the four-point correlation, i.e.,
\begin{align}
\langle \cR_{\vq_1}\cR_{\vk_1-\vq_1}\cR_{\vq_2}\cR_{\vk_2-\vq_2}\rangle =&
(2\pi)^3\,{\cal T}(\vq_1,\vk_1-\vq_1,\vq_2,\vk_2-\vq_2)\nn \\
& \times \delta^{(2)}(\vk_1+\vk_2)\,.
\end{align}
Regarding the form of the trispectrum ${\cal T}$, our analysis is motivated by inflationary models with a parity-violating Chern-Simons like coupling in the action, which can generate a parity-odd scalar trispectrum. 
The trispectrum of interest in these models are due to the exchange interaction constituted by
two three-point vertices and mediated by a gauge boson or 
graviton~\cite{Niu:2022fki,Fujita:2023inz,Reinhard:2024evr,Yura:2025mus,
Cho:2025rvg,Garcia-Saenz:2023zue,Creque-Sarbinowski:2023wmb,Moretti:2024fzb,Orlando:2025fec} 
(see also~\cite{Philcox:2025bvj} for a review of templates).
Thus, we define the parametric form of parity-even and parity-odd components of the scalar trispectrum 
arising from such exchange interactions as~\cite{Ragavendra:2025svk}

\begin{subequations}
\begin{align}
{\cal T}_{\rm even}(\vk_1,\vk_2,\vk_3,\vk_4) = & 2\,\taunl\, 
w_4(\vk_1,\vk_2,\vk_3,\vk_4)\nn\\
&\times P_\cR(k_1)P_\cR(k_3)P_\cR(\vert\vk_1+\vk_2\vert) \nn \\ 
&+\,\text{11 permutations}\,, \\
{\cal T}_{\rm odd}(\vk_1,\vk_2,\vk_3,\vk_4) =&\,i\,\tilde\tau_{\rm NL}\, 
w_4(\vk_1,\vk_2,\vk_3,\vk_4)\beta(\widehat{\vk_1+\vk_2}, \hvk_1, \hvk_3)\nn\\
&\times P_\cR(k_1)P_\cR(k_3) P_\cR(\vert\vk_1+\vk_2\vert) \nn\\
&+\,\text{23 permutations}\,,
\end{align}
\label{eq:trispec-paramet}
\end{subequations}
where the function $w_4$ captures the complete structure of the
trispectrum~[cf.~App.~\ref{app:3pt-4pt}].
The parameters $\tilde\tau_{\rm NL}$ and $\taunl$ quantify the
overall amplitude of the odd and even parts of the trispectrum respectively.
The function $\beta$ is the scalar triple product of the vectors in its arguments,
i.e., ~$\beta(\vk_1,\vk_2,\vk_3) = \vk_1 \cdot (\vk_2 \times \vk_3)$. 
It is explicitly parity-odd as it acquires negative sign under inversion of coordinates.
Also, we must note that ${\cal T}_{\rm odd}$ is purely imaginary while 
${\cal T}_{\rm even}$ is real, essentially to preserve the four-point correlation
in configuration space to be real.

Moreover, ${\cal T}_{\rm even}$ accounts for contributions to SIGWs
of order $\fnl^2\ps^3$ to SIGWs, the C-type and Z-type terms, mentioned earlier.
This is to say that the exchange trispectrum constructed above shall also include
trispectrum arising from scalar-exchanges, which shall be parity-even, along with
other possible contributions, such as from boson or graviton exchanges. 
We do not include the terms such as contact trispectrum or further higher-order 
correlations, quantified by $g_{_{\rm NL}}$ or $h_{_{\rm NL}}$, arising from 
higher-orders of perturbation theory in our analysis~\cite{Li:2023xtl,Perna:2024ehx}.

Having the necessary contributions to the four-point function at hand, we shall
write $\ph^I$ involving the power-, bi- and tri-spectra of the scalar perturbations 
as~\cite{Unal:2018yaa,Cai:2018dig,Adshead:2021hnm,Ragavendra:2021qdu,Perna:2024ehx,Ragavendra:2025svk}
\begin{widetext}
\begin{align}
\label{eq:ph-I-G}
\ph^I(k,\eta)\bigg\vert_{\rm Gaussian} =& 2\left(\f{8}{k^2\eta^2}\right)\f{k^3}{2\pi^2}
\int \f{\d^3\vq}{(2\pi)^3} 
\tilde I\left(\f{q}{k},\f{|\vk-\vq|}{k},\f{q}{k},\f{|\vk-\vq|}{k}\right)\,
\left[\vert Q^L(\vk,\vq)\vert^2+\vert Q^R(\vk,\vq)\vert^2\right]\,
P_\cR(q) P_\cR(|\vk-\vq|)\,, \\
\label{eq:ph-I-fnl}
\ph^I(k,\eta)\bigg\vert_{\rm hybrid} =& 8\left(\f{9}{25}\right)\left(\f{8}{k^2\eta^2}\right)
\f{k^3}{2\pi^2}\fnl^2 \iint\f{\d^3\vq_1\d^3\vq_2}{(2\pi)^6}
\tilde I\left(\f{q_1}{k},\f{|\vk-\vq_1|}{k},\f{q_1}{k},\f{|\vk-\vq_1|}{k}\right)
\left[\vert Q^L(\vk,\vq)\vert^2+\vert Q^R(\vk,\vq)\vert^2\right]\nn \\
&\times w_3^2(q_1,q_2,\vert\vq_1-\vq_2\vert)
P_\cR(|\vk-\vq_1|)P_\cR(q_2)P_\cR(|\vq_1-\vq_2|)\,,\\
\label{eq:ph-I-taunl}
\ph^I(k,\eta)\bigg\vert_{\rm C + Z} =& 4\left(\f{8}{k^2\eta^2}\right)
\f{k^3}{2\pi^2}\taunl \iint\f{\d^3\vq_1\d^3\vq_2}{(2\pi)^6}
\,\tilde I\left(\f{q_1}{k},\f{|\vk-\vq_1|}{k},\f{q_2}{k},\f{|\vk-\vq_2|}{k}\right)\nn \\
&\times \left[Q^L(\vk,\vq_1)Q^L(-\vk,-\vq_2)+ Q^R(\vk,\vq_1)Q^R(-\vk,-\vq_2)\right]\nn \\
&\times w_4(\vq_1,\vk-\vq_1,-\vq_2,\vq_2-\vk)
P_\cR(q_1)P_\cR(|\vq_1-\vq_2|)
\left[P_\cR(|\vk-\vq_1|)+P_\cR(|\vq_2-\vk|)\right]\,,
\end{align}
\end{widetext}
where we have defined the dimensionless function $\tilde I(u_1,v_1,u_2,v_2)$, 
following the time-averaging over fast oscillations and extracting out the overall 
time dependence $1/(k\eta)^2$ arising from the products of functions $I(q,|\vk-\vq|,\eta)$~[for explicit expressions, see App.~\ref{app:IQ}].

We have marked the various contributions to $\ph^I$ as Gaussian, hybrid, C-type and Z-type to 
relate them to the existing labels of these terms in the literature~\cite{Unal:2018yaa,Ragavendra:2021qdu,Perna:2024ehx}.
While $\ph^I\big|_{\rm Gaussian}$ is sourced by the power spectra of scalar
perturbations,
$\ph^I\big\vert_{\rm hybrid}$ is sourced by the cubic-order interaction 
giving rise to scalar bispectrum, and so is proportional to $\fnl^2$.
The term $\ph^I\big\vert_{\rm C + Z}$ is sourced by the even part of the
exchange trispectrum, and so is proportional to $\taunl$.
The total spectral density shall be $\ph^I = \ph^I\big\vert_{\rm Gaussian}
+ \ph^I\big\vert_{\rm hybrid}
+ \ph^I\big\vert_{\rm C + Z}$.
On the other hand, the power spectrum of V mode $\ph^V$ is simply
\begin{widetext}
\begin{align}
\label{eq:ph-V}
\ph^V(k,\eta) =& 4\left(\f{8}{k^2\eta^2}\right)
\f{k^3}{2\pi^2} \tilde\tau_{\rm NL}\iint\f{\d^3\vq_1\d^3\vq_2}{(2\pi)^6}
\,\tilde I\left(\f{q_1}{k},\f{|\vk-\vq_1|}{k},\f{q_2}{k},\f{|\vk-\vq_2|}{k}\right)
[Q^L(\vk,\vq_1)Q^L(-\vk,-\vq_2) - Q^R(\vk,\vq_1)Q^R(-\vk,-\vq_2)]\nn \\
&\times w_4(\vq_1,\vk-\vq_1,-\vq_2,\vq_2-\vk)
P_\cR(q_1)P_\cR(|\vq_1-\vq_2|) \left[P_\cR(|\vk-\vq_1|)+P_\cR(|\vq_2-\vk|)\right]\,.
\end{align}
\end{widetext}
As we can see, $\ph^V$ is sourced by the odd-part of the exchange trispectrum and so
is proportional to $\tilde{\tau}_{\rm NL}$. 
Lastly, we relate $\ph^I$ and $\ph^V$ to the associated spectral densities of SIGW
observed today as~\cite{Adshead:2021hnm} 

\begin{eqnarray}
\ogw^I(k) &=& \f{\Omega_{\rm rad}}{48}\left(\f{g_{\ast,0}}{g_{\ast,e}}\right)^{1/3}k^2\eta^2\ph^I(k,\eta)\,, \\
\ogw^V(k) &=& \f{\Omega_{\rm rad}}{48}\left(\f{g_{\ast,0}}{g_{\ast,e}}\right)^{1/3}k^2\eta^2\ph^V(k,\eta)\,,
\end{eqnarray}
where $\Omega_{\rm rad}$ is the fractional energy density of radiation today,
$g_{\ast,0}$ and $g_{\ast,e}$ are the effective number of relativistic degrees of freedom today
and at the end of inflation respectively.
We set their values to be $\Omega_{\rm rad}=4.2\times 10^{-5}$, $g_{\ast,e}=106.75$
and $g_{\ast,0}=3.36$~\cite{Planck:2018vyg}. 
Note that the combination $k^2\eta^2\ph^{I,V}(k,\eta)$ is a time independent quantity.

Recall that the non-Gaussianity parameters $\fnl$, $\taunl$ and $\tilde \tau_{\rm NL}$ can
have associated shapes and scale-dependencies (captured through $w_3$ and $w_4$) due to non-trivial 
primordial dynamics.
For our analysis, we are motivated by class of models that involve parity-violating interactions that 
lead to parity-odd trispectrum with scale dependence~\cite{Niu:2022fki,Fujita:2023inz,Reinhard:2024evr}. 

In such scenarios, it is motivated to choose two specific templates of scalar bispectrum namely, 
local and equilateral templates.
The templates in-turn determines the shape of the trispectrum as detailed in App.~\ref{app:3pt-4pt}.
Hence in our analysis, we shall work with local and equilateral templates of
scalar bispectrum and the associated shape of exchange trispectrum. 
While the functions $w_3=w_4=1$ for the local templates, they have non-trivial
structure for the equilateral template as given in Eqs.~\eqref{eq:w3-eq} and~\eqref{eq:w4_w3}.

To parametrize the enhancement of scalar power over small scales, while retaining the
behavior as constrained by CMB over large scales, we work with a lognormal
power spectrum $\ps(k)$, given by
\begin{align}
\ps(k) &= \As\left(\f{k}{k_\ast}\right)^{n_{\rm s} -1}
+ \f{A_{\rm p}}{\sqrt{2\pi\sigma^2_{\rm p}}}
\exp\left[-\f{1}{2\sigma_{\rm p}^2}
\ln^2\left( \f{k}{k_{\rm p}} \right)\right]\,.
\label{eq:ps-peak}
\end{align}
The parameters $\As$, $\ns$ are fixed to match the appropriate behavior over CMB scales with
$k_\ast=5\times 10^{-2}\,\mpcinv$.
We decide to fix the width of the peak $\sigma^2_{\rm p} = 0.1$ and to vary the value of the peak amplitude of the 
spectrum $A_{\rm p}$, and the location of the peak $k_{\rm p}$. 
In our forecasts, we also include the non-Gaussianity parameters $\fnl$, $\taunl$ and $\tilde \tau_{\rm NL}$
as model parameters. We shall work with peak located at $k_{\rm p}=10^{16}-10^{17}\,\mpcinv$ to obtain the resultant 
$\ogw^{I,V}$ in the frequency bands of ET and CE.

Using the explicit form of $\ps(k)$, we numerically compute the associated $\ogw^{I,V}$.
Upon obtaining the spectral densities, we establish the following scaling relations in terms 
of the underlying model parameters.
\begin{align}
    \Omega_{\rm GW}^I(f) &= \f{\Omega_{\rm rad}}{48}\left(\f{g_{\ast,0}}{g_{\ast,e}}\right)^{1/3}\,\Big[ A_{\rm p}^2\,{\cal I}_I^{(1)}(k,k_{\rm p}, \sigma_{\rm p}) \notag  \\
    &+  f_{\rm NL}^2 \,  A_{\rm p}^3\, \mathcal{I}_I^{(2)} (k, k_{\rm p}, \sigma_{\rm p})\notag \\
    & + \tau_{\rm NL}\,A_{\rm p}^3\,{\cal I}_I^{(3)}(k,k_{\rm p}, \sigma_{\rm p}) \Big]\,, \label{eq:full_spectra_I} \\ 
    \Omega_{\rm GW}^V(f) &= \f{\Omega_{\rm rad}}{48}\left(\f{g_{\ast,0}}{g_{\ast,e}}\right)^{1/3}\, \tilde{\tau}_{\rm NL}\,A_{\rm p}^3\,{\cal I}_V^{(3)}(k,k_{\rm p}, \sigma_{\rm p})\,,
    \label{eq:full_spectra_V}
\end{align}
where the factors ${\cal I}_{I,V}$ are obtained by numerical computation of integrals involved 
in Eqs.~\eqref{eq:ph-I-G}, \eqref{eq:ph-I-fnl}, \eqref{eq:ph-I-taunl} and~\eqref{eq:ph-V} after extraction of overall dependences on parameters $A_{\rm p}$, $\fnl$, $\taunl$ and 
$\tilde \tau_{\rm NL}$~\cite{Ragavendra:2025svk}. 
These relations allow us to separate the shapes of $\ogw^{I,V}$, written in terms of 
the numerical functions ${\cal I}_{I,V}$, from their dependence on the parameters of interest, 
namely $A_{\rm p}$, $f_{\rm p}$, $\fnl$, $\taunl$ and $\tilde \tau_{\rm NL}$.
Note that the spectral densities $\ogw^{I,V}$ are expressed in terms of frequencies $f$ 
using the linear relation between $f$ and $k$
\begin{align}
\f{f}{\rm Hz} =& 1.55\left(\f{k}{10^{15}\,\mpcinv}\right).
\end{align}

In our analysis we consider also the presence of an isotropic astrophysical background that can arise 
from unresolved compact binaries \cite{Phinney:2001di}, and that can act as a foreground for our cosmological signal.
The spectral density of such an astrophysical background is modeled as a power-law in frequency as
\begin{align}
\ogw^{\rm astro} =& A\,\left( \frac{f}{25 \, \rm Hz}\right)^{n}\,.
\label{eq:ogw-astro}
\end{align}
This contribution can affect the detectability of the intensity, and we will consider it in our analysis. For the chiral contribution, there could be a similar foreground, generated by the Poisson fluctuations in the number of unresolved sources~\cite{ValbusaDallArmi:2023ydl}, that we disregard in our analysis. 
For the astrophysical foreground, we assume the fiducial values of $n=2/3$, $\log_{10}A=-9$, 
in agreement with LVK O4a upper bounds~\cite{LIGOScientific:2025bgj}.

\begin{figure}
\includegraphics[width=\linewidth]{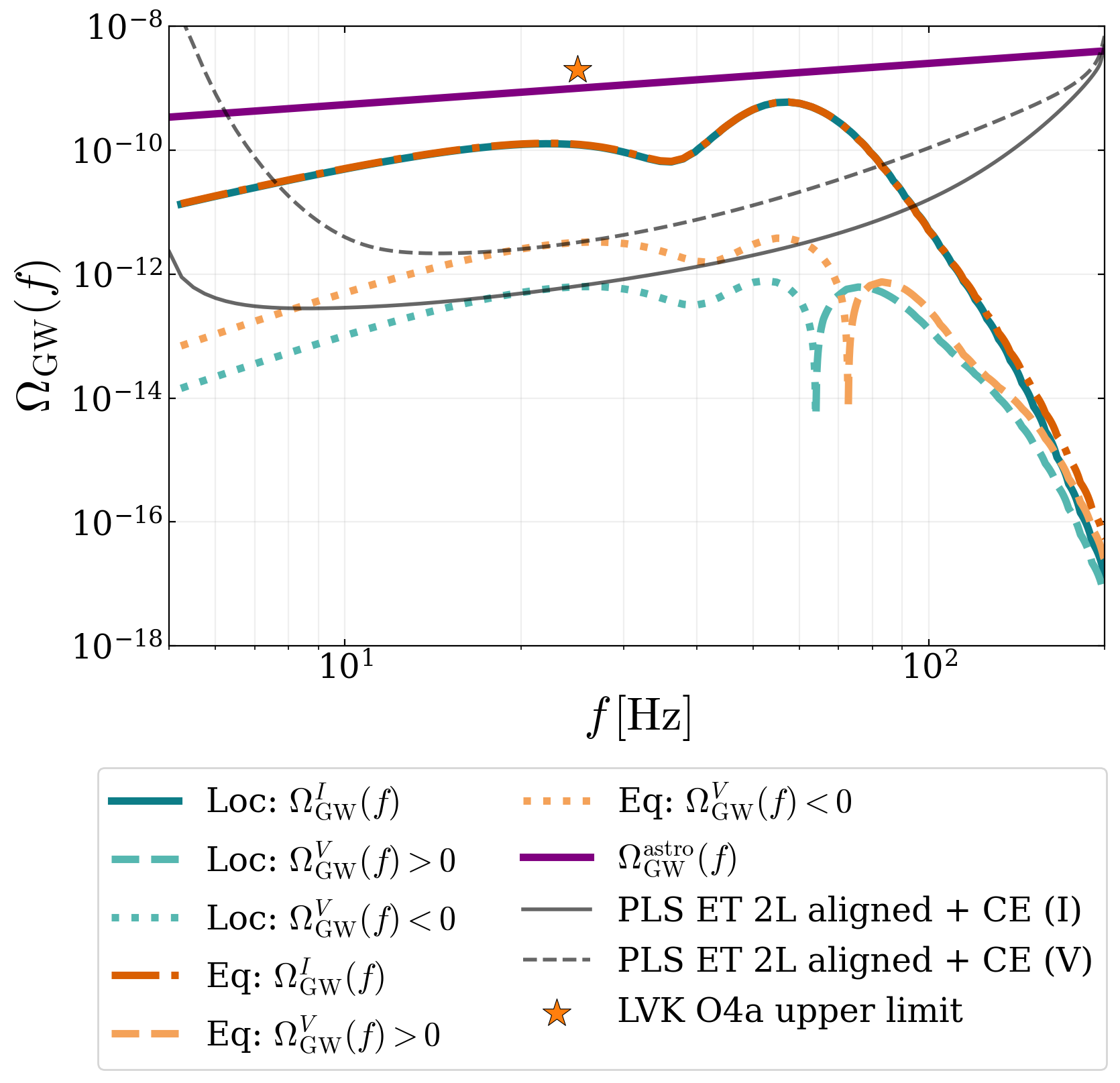}
\caption{Plot of the SIGW spectral densities for the $I$ and $V$ modes,
obtained using the local and equilateral  templates of bispectrum and corresponding trispectrum, with the following parameters:
$A_{\rm p}=10^{-2},\,f_{\rm peak}=50\,{\rm Hz},\,\sigma_p^2 = 10^{-1},\,
\fnl = 5, \taunl=10$ and $\tilde{\tau}_{\rm NL}=10$. 
We also plot the astrophysical background with corresponding parameters set to $\log_{10}A=-9$ and $n=2/3$ (cf.~Eq.~\eqref{eq:ogw-astro}), 
motivated by the upper bound by the LVK collaboration. 
Power Law Integrated Sensitivity curves (PLS), evaluated with \cite{Caporali:2025dyf}, are displayed in gray for the ET 2L + CE network, considering $T_{\rm obs}= 1\, \rm yr$ and $\rm SNR_{thr}=1$, both for Intensity (solid) and chiral GWs (dashed).
We consider the shapes of $\ogw^I$ and $\ogw^V$ and the associated parametric dependencies (cf. Eqs.~\eqref{eq:full_spectra_I} and~\eqref{eq:full_spectra_V}) to perform Bayesian parameter estimation considering ET in the 2L aligned configuration and CE.}
\label{fig:ogw-loc}
\end{figure}

For representative values of model parameters, we show the spectral densities of SIGW along with the astrophysical component in Figure ~\ref{fig:ogw-loc}.
Considering the two different shapes in $\ogw^I$ and $\ogw^V$ and the associated parametric
dependencies on $A_{\rm p}$, $\fnl$, $\taunl$, $\tilde \tau_{\rm NL}$,
we simulate the corresponding data that may be seen in the ET+CE network, and we forecast the 
bounds that can be put on such parameters, using third-generation detectors.

\section{Bayesian Analysis for the SGWB}
\label{sec:Bayesian Analysis for the SGWB}

In this section, we summarize the main points for performing a Bayesian analysis of the model considered.

\subsection{Data generation}
\label{sec:Datagen}

In principle, the idea is to generate the data streams that will contain the SGWB signal that one aims to estimate, along with the noise.
In the case of ground-based detectors, we can sketch the output data $d_i(t)$ in the detector $i$ as 

\begin{equation}
    d_i(t) = h_i(t) + n_i(t) \, ,
    \label{eq:data_time_domain}
\end{equation}
where $h_i(t)$ and $n_i(t)$ are the time-domain signal and noise contributions, respectively.

Assuming that every transient signal has been removed, the data have zero mean and covariance given by 
\begin{equation}
    {\rm cov}_{ij}(t,t^\prime) \equiv \left\langle d_i(t)d_j(t^\prime)\right\rangle \, .
    \label{eq:covariance_data_f}
\end{equation}

Assuming a stationary signal and noise, we perform the analysis in the frequency domain, working directly with the Fourier-transformed data.
Therefore 
\begin{align}
     \langle \tilde{d}_i (f) \, \tilde{d}_j^*(f)\rangle \, &=    \langle (\tilde{h}_i(f) + \tilde{n}_i(f)) \, \, (\tilde{h}_j^*(f) + \tilde{n}_j^*(f))\rangle \nonumber \\
     &= \langle \tilde{h}_i(f) \, \tilde{h}_j^*(f)\rangle \, + \langle \tilde{n}_i(f) \, \tilde{n}_j^*(f)\rangle\,,
    \label{eq:covariance_data}
\end{align}

where we assume that the noise and the signal are uncorrelated, i.e., $\langle \tilde{h}_{i,j} (f)\, \tilde{n}_{i, j}(f)\rangle=0$.

At this point, for the signal part, we can consider the time-averaged estimator, defined as \cite{LIGOScientific:2019vic} \cite{Romano:2016dpx}

\begin{equation}
    \hat{C}_{ij}(f) \equiv \frac{1}{N_{seg}}\sum_t\frac{2}{T_{\rm{seg}} S_{0}(f)}\Re[\tilde{d}_{i}(t,f)\tilde{d}^{*}_{j}(t,f)]\,, 
    \label{eq:hatC}
\end{equation}

where the sum runs over the time segments and $T_{\rm seg}= 4\, \rm s$, $N_{\rm seg}\equiv T_{\rm obs}/T_{\rm seg}$ the number of time segments employed in the analysis and $T_{\rm obs}$ the observation time, with the factor $S_{0}(f) \equiv 3H_{0}^{2}/10 \pi^{2} f^{3}$ used to match the units of the square of the data to those of the spectral energy density of the SGWB.
Averaging over many time segments allows us to treat the estimator of the SGWB as a Gaussian random variable, due to the central limit theorem \cite{Romano:2016dpx}. 
We can therefore use Eq. \eqref{eq:hatC} to directly generate the signal part in Eq. \eqref{eq:covariance_data} and Eq. \eqref{def:PSD_noise} for the noise.

\subsection{Likelihood}
\label{subsec:likelihood}

The time-average of the estimator, defined in  Eq. \eqref{eq:hatC}, is 
\begin{equation}
    \bar{C}_{ij}(f) \equiv \langle \hat{C}_{ij}(f)\rangle = \gamma_{ij}(f)\Omega_{\rm GW}(f) + \frac{10 \pi f^3}{3H_0^2}N_{ij}(f),
    \label{eq:barCIJ}
\end{equation}
where $\gamma_{ij}$ is the normalized ORF.  Then, the full expression reads 
\begin{equation}
\bar{C}_{ij}(f) = \gamma_{ij}^t(f)\Omega^I_{\rm GW}(f) + \gamma_{ij}^V(f)\Omega^V_{\rm GW}(f) + \frac{10 \pi f^3}{3H_0^2}N_{ij}(f) \, .
\end{equation}
 For $i=j$, since $\gamma_{ii}^t= 1$ and $\gamma_{ii}^V= 0$
\begin{equation}
\bar{C}_{ii}(f) = \Omega^I_{\rm GW}(f) +  \frac{10 \pi f^3}{3H_0^2}N_{ii}(f) \, , 
\end{equation}
where $N_{ii}(f)$ is the PSD of the detector. While, for $i \neq j$
\begin{equation}
\bar{C}_{ij}(f) = \gamma_{ij}^t(f)\Omega^I_{\rm GW}(f) + \gamma_{ij}^V(f)\Omega^V_{\rm GW}(f) \, , 
\end{equation}
since we do not consider correlated noise~\cite{Caporali:2025mum}.

In~\cite{Romano:2016dpx} it has been shown that the variance of the estimator is proportional to the SGWB signal itself (intrinsic variance) and to the PSD of the noise,
\begin{equation}
    \Sigma_{ij}(f) \equiv \left\langle \left[\hat{C}_{ij}(f)-\bar{C}_{ij}(f)\right]^2 \right\rangle = \frac{\bar{C}_{ii}\bar{C}_{jj}+\bar{C}_{ij}^2}{2 N_{\textrm{seg}}} \, . 
    \label{eq:SigmaIJ}
\end{equation}

The full likelihood of the data incorporates information from all $i$ and $j$ channels in the detector network.

Given that $\gamma_{ii}^V=0$ and $N_{ij}=0$ for $i \neq j$,
\begin{align}
    \Sigma_{ij} = \frac{(\Omega_{\rm GW}^I + \frac{N_{ii}}{S_0}) \, (\Omega_{\rm GW}^I + \frac{N_{jj}}{S_0}) + (\gamma_{ij}^t \Omega_{\rm GW}^I + \gamma_{ij}^V \Omega_{\rm GW}^V)^2}{2 N_{seg}}
\end{align}
where the dependency on the frequency of the considered quantities is implicit.

For the 2L configuration (here and thereafter, we are going to refer to the two interferometers as ET L1 and ET L2, respectively placed in the two proposed candidate sites of Sardinia and the Netherlands \cite{Branchesi:2023mws, ET:2025xjr}, in the so-called ``aligned'' configuration) and CE, we consider only the cross-correlation channels between all the detector pairs, as we assume that the auto-correlations are used to reconstruct the PSD. Therefore, we fix the PSD to the reference value provided in~\cite{Branchesi:2023mws} for ET, as well as for CE{\footnote{https://dcc.ligo.org/LIGO-T1500293/public}}.

We can thus write down a Gaussian likelihood 
\begin{equation}
    \mathcal{L} = \prod_{i,j} \frac{1}{\sqrt{2\pi \Sigma_{ij}}} {\rm exp}\left[-\frac{1}{2}\frac{(\hat{C}_{ij}-\bar{C}_{ij})^2}{\Sigma_{ij}}\right] \, , 
    \label{eq:likelihood_estimator}
\end{equation}

where we consider the pairs $(i,j)=\{ \rm (ET \, L1, ET \, L2), (ET \, L1, CE)\}, (ET \, L2, CE)$.

\subsection{Parameters Choice and Priors discussion}
\label{sec:Priors}

\paragraph{Choice of the injected values}

Since our goal is to probe scalar non-Gaussianity parameters that enter in the SIGW energy density budget, we impose a physically motivated hierarchy among the different contributions entering in Eq.~\eqref{eq:full_spectra_I}. 

The total intensity spectrum of cosmological origin can be schematically decomposed as 
\begin{equation}
\Omega_{\rm GW }^{I}(f)
=
\Omega^{\mathcal{P}}_{\rm GW}(f)
+
\Omega^{\mathcal{B}}_{\rm GW}(f)
+
\Omega^{\mathcal{T}}_{\rm GW}(f)\,,
\end{equation}
where, comparing with Eq.~\eqref{eq:full_spectra_I}, we may write
\begin{align}
\Omega^{\mathcal{P}}_{\rm GW}(f) &= \f{\Omega_{\rm rad}}{48}\left(\f{g_{\ast,0}}{g_{\ast,e}}\right)^{1/3}\,A_{\rm p}^2\,{\cal I}_I^{(1)}(k,k_{\rm p}, \sigma_{\rm p})\,,\\
\Omega^{\mathcal{B}}_{\rm GW}(f) &= \f{\Omega_{\rm rad}}{48}\left(\f{g_{\ast,0}}{g_{\ast,e}}\right)^{1/3}\,
f_{\rm NL}^2 \, A_{\rm p}^3\, \mathcal{I}_I^{(2)} (k, k_{\rm p}, \sigma_{\rm p})\,, \label{eq:bisOmega}\\
\Omega^{\mathcal{T}}_{\rm GW}(f) &= \f{\Omega_{\rm rad}}{48}\left(\f{g_{\ast,0}}{g_{\ast,e}}\right)^{1/3}\,
\tau_{\rm NL}\,A_{\rm p}^3\,{\cal I}_I^{(3)}(k,k_{\rm p}, \sigma_{\rm p})\,.
\end{align}

where the factors of ${\cal I}_I$ are numerical functions determining the shape of the spectra as mentioned before.

While choosing the injection values, we ensure preserving the perturbativity of the scalar perturbations,
in that we impose the conditions of $\fnl^2 A_{\rm p}^3 < A_{\rm p}^2$ and $\taunl A_{\rm p}^3 < A_{\rm p}^2$.
Thus we have the hierarchy on the combination of parameters as
\begin{align}
  &  f_{\rm NL}^2 \, A_{\rm p} < 1 \,, \nonumber\\
  & \tau_{\rm NL} \, A_{\rm p} < 1 \,. 
\label{eq:hierarchy_condition}
\end{align}

Concerning the parity-violating signal encoded in the circular polarization spectrum
\begin{align}
\Omega_{\rm GW}^{V}(f) =&\f{\Omega_{\rm rad}}{48}\left(\f{g_{\ast,0}}{g_{\ast,e}}\right)^{1/3}\,
\tilde \tau_{\rm NL}\,A_{\rm p}^3\,{\cal I}_V^{(2)}(k,k_{\rm p}, \sigma_{\rm p})\,,
\end{align}

The physical consistency of the Stokes parameters requires that the polarized component does not exceed the total intensity, i.e.,
\begin{equation}
\left| \frac{\tilde{\tau}_{\rm NL} A_{\rm p}}{ 1 +  f_{\rm NL}^2 A_{\rm p}  +  \tau_{\rm NL} A_{\rm p} } \right| < 1
\label{eq:stokes_condition}
\end{equation}
ensuring that the degree of circular polarization remains smaller than the contribution of the intensity. 
However, since the intensity and chirality over the UV tail is dominated by non-Gaussian contributions, the Stokes relation over this regime becomes
\begin{equation}
    \tilde \tau_{\rm NL} \leq f_{\rm NL}^2 + \tau_{\rm NL}.
    \label{eq:stricter_stokes}
\end{equation}
Eq.~\eqref{eq:stricter_stokes} is a stricter condition than Eq.~\eqref{eq:stokes_condition}.
If this condition is satisfied, then Eq.~\eqref{eq:stokes_condition} is also satisfied, 
because $\fnl^2 A_p < 1$ and $\tau_{\rm NL} A_p < 1$.
Moreover, the latter condition ensures the Stokes inequality over both the peak and the tail, 
while the former ensures it only at the peak.

We enforce the conditions given in Eqs.~\eqref{eq:hierarchy_condition}--\eqref{eq:stokes_condition}
directly at the level of the GW spectrum, so that the choice of fiducial parameters is physically motivated, and we also ensure that they are within the constraints given 
by~\cite{Philcox:2025wts,Planck:2019kim}.

\paragraph{Choice of the priors}

We adopt physically uninformative priors for the cosmological parameters associated with the higher-order non-Gaussian corrections to the induced GW spectrum. In particular, we do not impose the restrictions given in Eqs.~\eqref{eq:hierarchy_condition}--\eqref{eq:stokes_condition}. This choice prevents the priors from artificially shaping the posterior distributions, which could otherwise bias our parameter estimates. 

Moreover, since $f_{\rm NL}$ enters quadratically in Eq. \ref{eq:bisOmega}, it is not possible to determine its intrisic sign. Therefore, to avoid multimodal symmetical posteriors, we use a uniform positive prior for $f_{\rm NL}$. The results we obtain has to be taken as an estimate of its absolute value, since from the model we are not able to distinguish its sign.

The priors used in the analysis are summarized in Table \ref{tab:priors}.

\begin{table}[!t]
    \centering
    \begin{tabular}{c|c}
        \hline
        \hline
         Parameter & Prior \\
         \hline
         \multicolumn{2}{c}{Astrophysical parameters} \\
         \hline
          $\log_{10}A$ & $\mathcal{U}[-15,\, -6]$ \\
          $n$ & $\mathcal{U}[-5,\, 5]$ \\
          \hline
        \multicolumn{2}{c}{Cosmological parameters (I)} \\
        \hline
        $f_{\rm peak}$ & $\mathcal{U}[1,\, 400]$ \\
        $A_p$ & $\log \mathcal{U}[10^{-4},\, 1]$ \\
        $f_{\rm NL}$   & $\mathcal{U}[0,\, 100]$ \footnote{The choice of having only positive values comes from the fact that $f_{\rm NL}$ enters quadratically in Eq.\eqref{eq:full_spectra_I}, therefore, the choice of a symmetric prior would lead to a symmetric bimodal distribution for the posterior of $f_{\rm NL}$. In this sense, we are only able to constrain the absolute value of $f_{\rm NL}$, not its sign.} \\
        $\tau_{\rm NL}$   & $\mathcal{U}[-10^4,\, 10^4]$ \\
        \hline
        \multicolumn{2}{c}{Cosmological parameters (V)} \\
        \hline
        $\tilde{\tau}_{\rm NL}$   & $\mathcal{U}[-10^4,\, 10^4]$ \\
        \hline
    \end{tabular}
    \caption{Prior distributions used in the analysis.
    Uniform priors are denoted by $\mathcal{U}[a,b]$,
    log-uniform priors by $\log\mathcal{U}[a,b]$.}
    \label{tab:priors}
\end{table}

\section{Results}
\label{sec:Results}

\subsection{Parameter Estimation}
\label{subsec:PE}

\begin{table}[]
    \centering
    \begin{tabular}{c|c| c}
    \hline
    \hline
     & \multicolumn{2}{c}{SNR} \\
     \hline 
     & Local & Equilateral \\
     \hline
         $\Omega_{\rm GW}^{\rm astro}$ & \multicolumn{2}{c}{966} \\
         \hline
         $\Omega_{\rm GW}^{\mathcal{P}}$ & \multicolumn{2}{c}{103} \\
         \hline
         $\Omega_{\rm GW}^{\mathcal{B}}$ & 2.3 & 1.3\\
         $\Omega_{\rm GW}^{\mathcal{T}}$ & 4.7 & 15 \\
         $\Omega_{\rm GW}^{\rm V}$ & 0.5 & 1.9\\
        \hline
    \end{tabular}

    \caption{SNR values for the different contributions to the spectra, 
    in the cases of the local and equilateral templates, 
    evaluated with \cite{Caporali:2025dyf}.
    The observation time considered is $T_{\rm obs}= 1\, \rm yr$. }
    \label{tab:SNR}
\end{table}

\begin{figure*}
    \centering
    \includegraphics[width=1\linewidth]{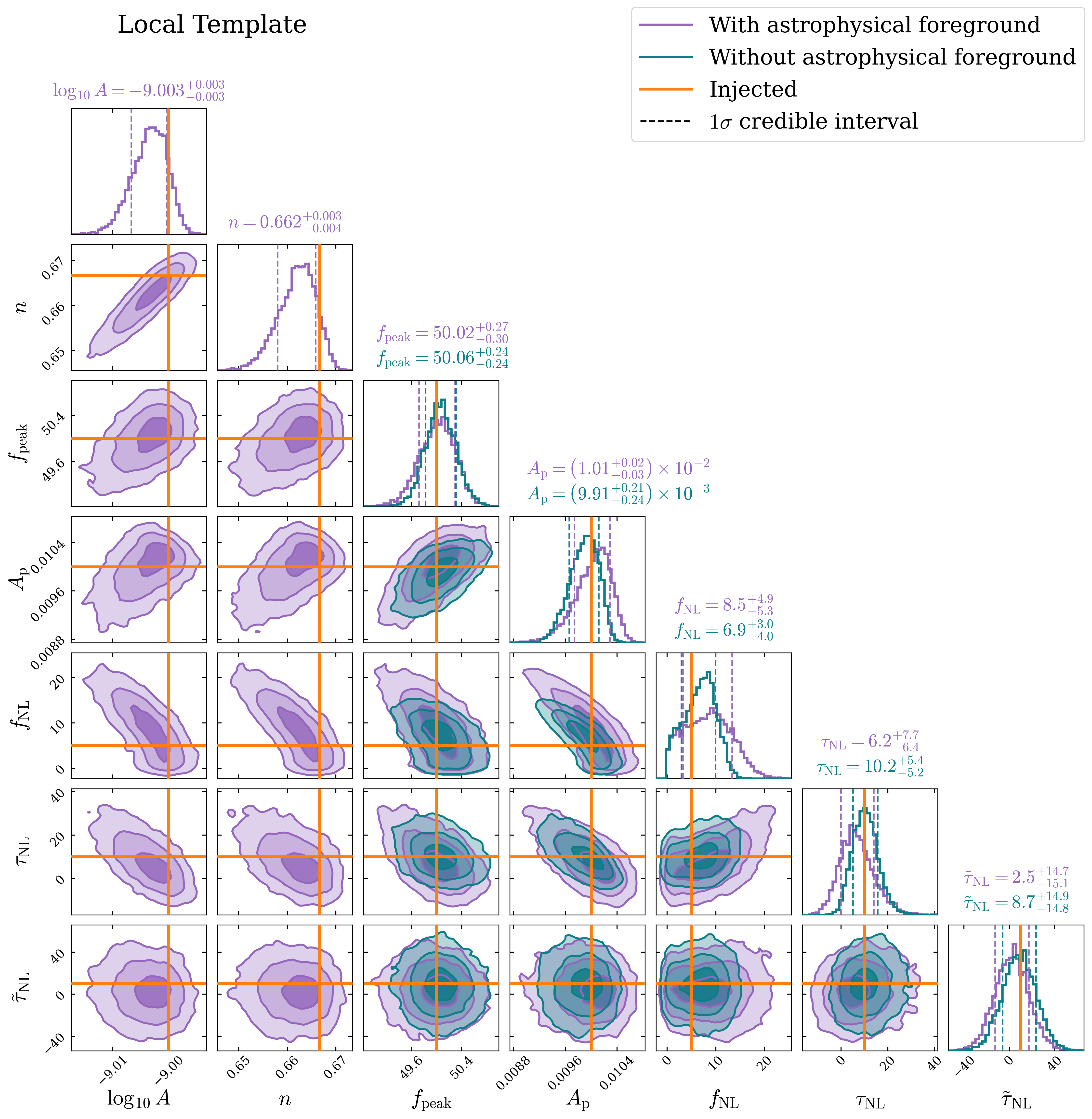}
    \caption{Parameter estimation for the case of local template
    with $\sigma_p^2 = 10^{-1}$, assuming an observation time $T_{\rm obs}=1\,{\rm yr}$. 
Purple and teal contours show the posterior distributions obtained with and without the astrophysical foreground, respectively. 
Orange lines mark the injected values, while dashed lines in the one-dimensional marginalized posteriors indicate the $1\sigma$ credible intervals.}
    \label{fig:corner_loc}
\end{figure*}

\begin{figure*}
    \centering
    \includegraphics[width=1\linewidth]{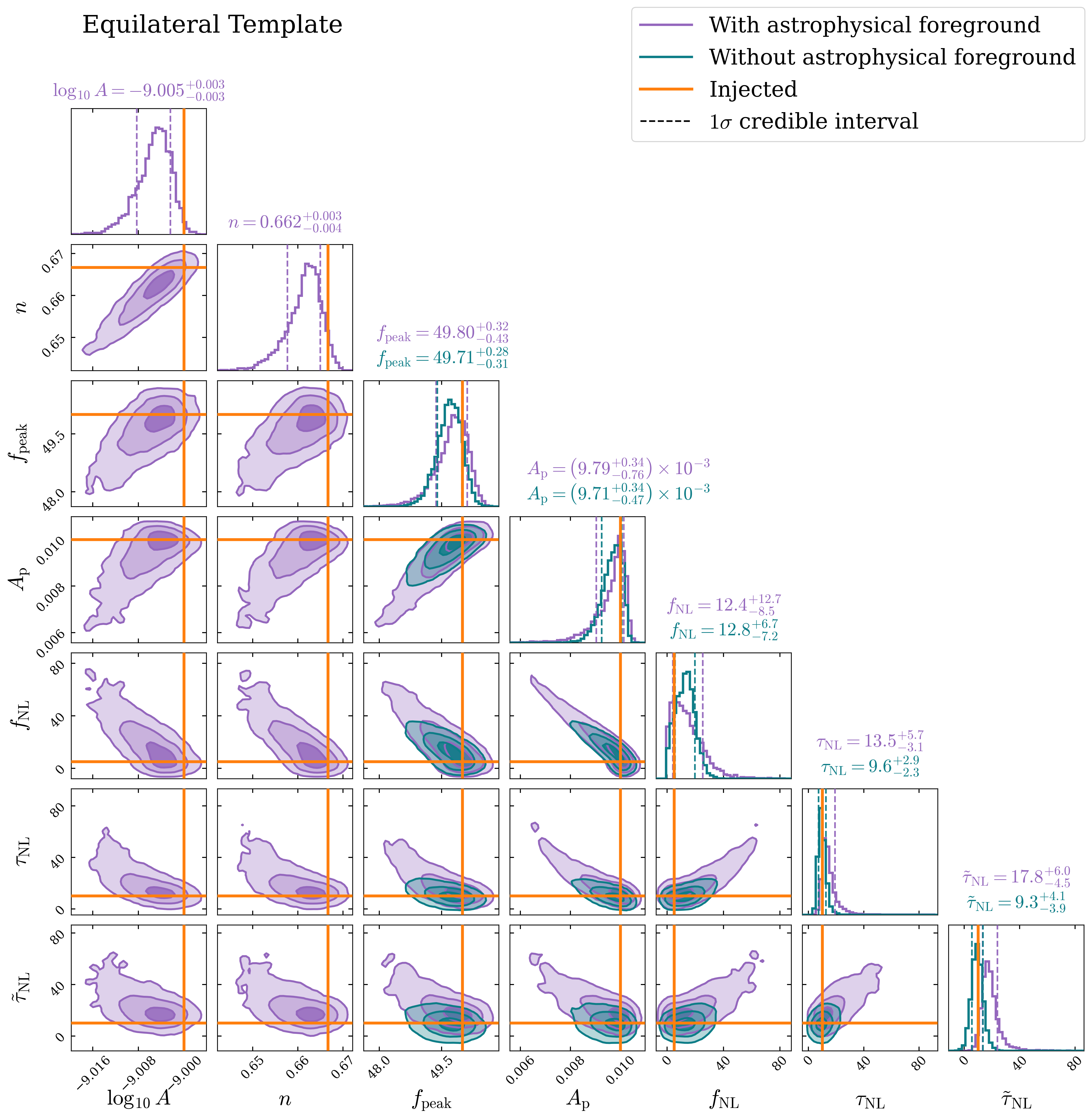}
    \caption{Parameter estimation for the the case of equilateral template 
    with $\sigma_p^2 = 10^{-1}$, assuming an observation time $T_{\rm obs}=1\,{\rm yr}$. 
Purple and teal contours show the posterior distributions obtained with and without the astrophysical foreground, respectively. 
Orange lines mark the injected values, while dashed lines in the one-dimensional marginalized posteriors indicate the $1\sigma$ credible intervals.}
    \label{fig:corner_eq}
\end{figure*}

We perform a Bayesian parameter estimation by implementing the likelihood presented in Eq.~\eqref{eq:likelihood_estimator} in \texttt{Bilby} \cite{Ashton:2018jfp,Romero-Shaw:2020owr,Morisaki:2023kuq} using the \texttt{Dynesty} sampler \cite{Speagle:2019ivv}. The choice of the injected values of the parameters is described in Section \ref{sec:Priors}, which are $f_{\rm peak}=50 \, {\rm Hz}, \,  A_p=10^{-2},\, \fnl = 5, \taunl=10, \tilde{\tau}_{\rm NL}=10 $. 
We generate the data as described in Section \ref{sec:Datagen} for $T_{\rm obs} = 1 \, \rm  yr$ with $T_{\rm seg}=4 \, \rm s$. Data are generated in the frequency domain from $f_{\rm min} = 5 \, \rm  Hz$ to $f_{\rm max}=200 \, \rm Hz$ with a resolution of $\Delta f = 1/ T_{\rm seg}=0.25$.
In Table \ref{tab:SNR}, we show the Signal-to-noise ratios (SNR) of the different contributions to the spectra, both in the case of the local and equilateral non-Gaussianity templates considered.

In Figure \ref{fig:corner_loc} and Figure \ref{fig:corner_eq}, we present the posterior distributions for the parameters respectively in the case of local and equilateral templates of bispectra and associated trispectra. We compare both templates with (in purple) and without  (in teal) an astrophysical foreground contribution.

Considering the case of local template in Figure \ref{fig:corner_loc}, the frequency peak \(f_{\rm peak}\) and amplitude of the signal \(A_p\) are well recovered within $1 \sigma$, with the orange lines lying within the main credible regions for all parameters. The inclusion of the astrophysical foreground does not induce a significant bias in the reconstruction of the primordial-spectrum parameters, whose posteriors remain centered close to the injected values. 
Concerning the marginalized posteriors for the non-Gaussianity parameters $f_{\rm NL}$ and ${\tau}_{\rm NL}$, they are wider but remain broadly consistent with the injected values.
They are all reconstructed within  $1 \sigma$, both when the astrophysical foreground is included and when it is not.
Concerning $\tilde{\tau}_{\rm NL}$, even though the posterior distribution is compatible with zero, 
the reconstruction of the injected value occurs within $1\sigma$. 
 
The marginalized distribution of $\tilde{\tau}_{\rm NL}$ shows less accuracy
than the other parameters, and it can be attributed to the lower
${\rm SNR}=0.5$ for $\ogw^V$ (cf.~Table~\ref{tab:SNR}).

The corner plot also highlights several correlations among parameters. Among the primordial parameters, $A_p$ and $f_{\rm peak}$ show a visible positive correlation. This behavior is expected since the chosen peak in frequency of the signal happens at higher frequencies than the ones at which both ET and CE are most sensitive to. Looking instead at the correlation between $A_{\rm p}$ and $f_{\rm NL}$, this almost banana-shaped contour can be due to the dependence of $\ogw^{\cal B} \propto \fnl^2\,A_{\rm p}^3$ [cf.~Eq.~\eqref{eq:ph-I-fnl}]. In particular, larger non-Gaussian contributions can be partially compensated by changes in the amplitude and peak position of the induced GW spectrum. These degeneracies become more pronounced when the astrophysical foreground is included. 

For the astrophysical foreground parameters $\log_{10}A$ and $n$, they both show good reconstruction within $2\sigma$ and a very small relative error, with a clear correlation 
between \(\log_{10}A\) and \(n\), reflecting the expected degeneracy between amplitude-tilt parametrization. This is because the contribution of the foreground to the total spectrum carries an SNR=966, which is 1 to 2 orders of magnitude larger than the various components of the cosmological signal 
(cf.~Table~\ref{tab:SNR}).  
In principle, the foreground could not only degrade the reconstruction of the cosmological parameters, but also potentially induce a structured bias in the inferred parameters by partially absorbing or reshaping the signal. 
However, we notice that the main effect of including the astrophysical foreground is a broadening of the posterior distributions for the non-Gaussian parameters. 
Despite this effect, all the cosmological parameters are still reconstructed within $1\sigma$ for the local template. Concerning the astrophysical foreground parameters, they are reconstructed within $2\sigma$.
This shows that the developed pipeline is robust and well-suited for taking into account several parameters of different origins, including non-Gaussianity and chirality.

In Figure \ref{fig:corner_eq}, we plot the posterior distributions for the equilateral template, 
which behave in a similar way to the local template, both in the case with and without the astrophysical foreground. 
However, we see that $f_{\rm NL}$ in this template shows wider posteriors.
This can be attributed to $\Omega_{\rm GW}^{\mathcal{B}}$ contribution having a lower SNR value (Table~\ref{tab:SNR}) in the equilateral template with respect to the local one. 
On the other hand,the same line of reasoning can be applied to the $\tau_{\rm NL}$ and $\tilde{\tau}_{\rm NL}$ estimates: the $\Omega_{\rm GW}^{\mathcal{T}}$ and $\Omega_{\rm GW}^{V}$ present larger SNR value in the case of the equilateral template compared to the local one, showing a higher constraining power in the parameter estimation. Thus it allows for more precise constraints on these two parameters in the equilateral template
compared to the local template. 
It is worth mentioning that even in the equilateral template, 
the addition of the astrophysical foreground leads to the broadening of the posteriors. Despite this, we are still able to constrain within $2\sigma$ all the parameters involved in our analysis.

\subsection{Polarization Degree constraints}
\label{subsec:Pi}
We may quantify the relative degree of circular polarization in a given GW signal 
as~\cite{Satoh:2010ep,Orlando:2020oko,Martinovic:2021hzy,Duval:2025vfg}

\begin{align}
    \Pi (f) &= \frac{\Omega_{\rm GW}^{V}(f)}{\Omega_{\rm GW}^{I}(f)}\,, 
    \label{eq:Pi}
\end{align}
which, by the Stokes inequality of 
$\Omega_{\rm GW}^I(f) \geq \Omega_{\rm GW}^{V} (f) \, \, \, \forall f$ is bounded to be $\Pi \in [-1, 1]$. 

In case of SIGWs, $\Pi(f)$ can be recast in terms of individual spectral contributions as
\begin{align}
    \Pi (f) &= \frac{\Omega_{\rm GW}^{V}(f)}{\Omega_{\rm GW}^{I (\mathcal{P})}(f)+ \Omega_{\rm GW}^{I (\mathcal{B})}(f)+ \Omega_{\rm GW}^{I (\mathcal{T})}(f)}\,,\nonumber \\
    & = \frac{\tilde{\tau}_{\rm NL} A_p \mathcal{I}^{(2)}_V}{ \mathcal{I}^{(1)}_I + f_{\rm NL} A_p \mathcal{I}^{(3)}_I + \tau_{\rm NL} A_p \mathcal{I}^{(2)}_I}\,, 
\end{align}
where the dependence over $(k, k_{\rm peak}, \sigma_p)$, i.e., ($f, f_{\rm peak}$) is implicit in the integration factors ${\cal I}$. 

\begin{figure}
    \centering
    \includegraphics[width=1\linewidth]{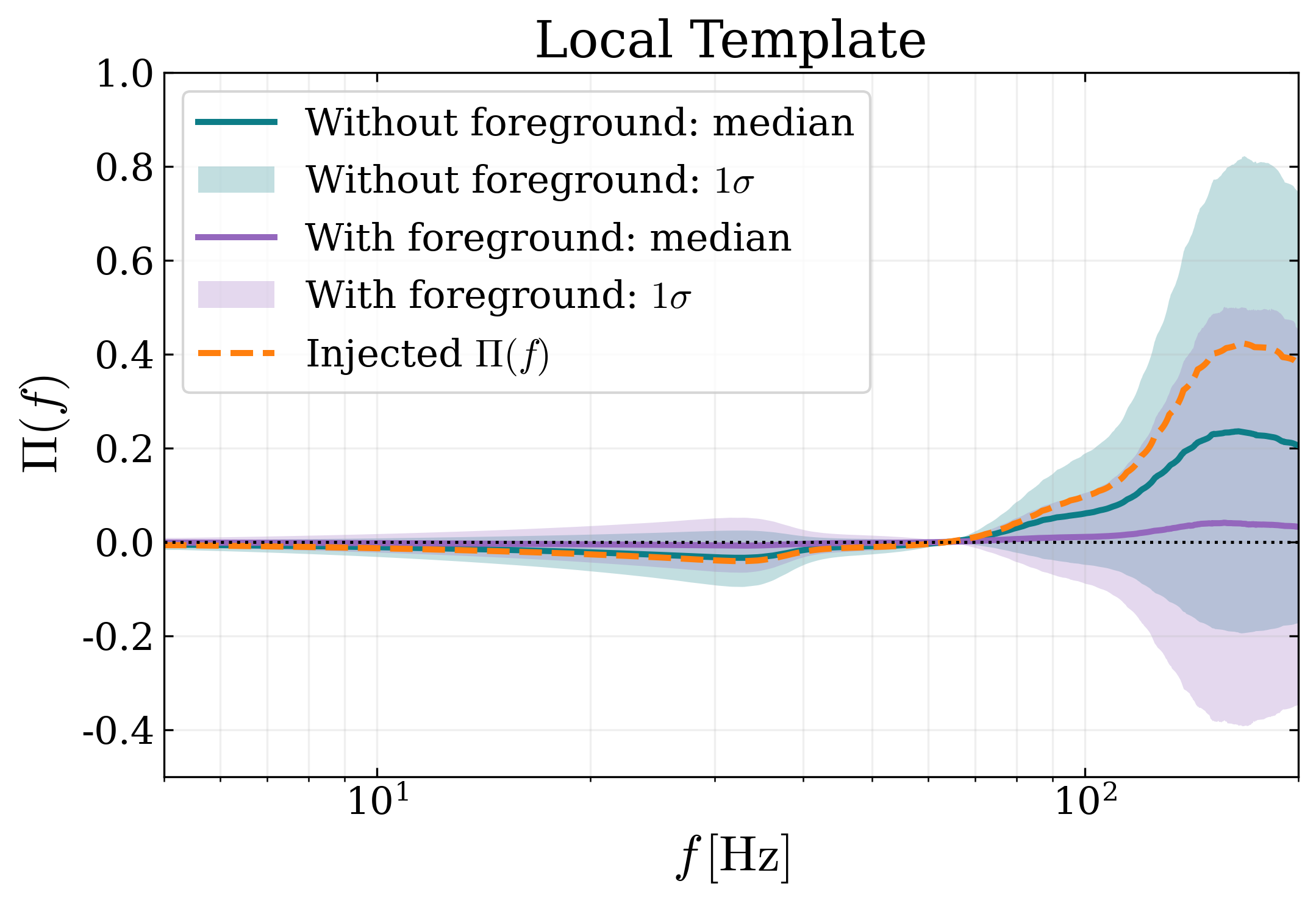}
    \caption{Reconstructed degree of polarization for the local template.} 
    \label{fig:Pi_loc}
\end{figure}

\begin{figure}
    \centering
    \includegraphics[width=1\linewidth]{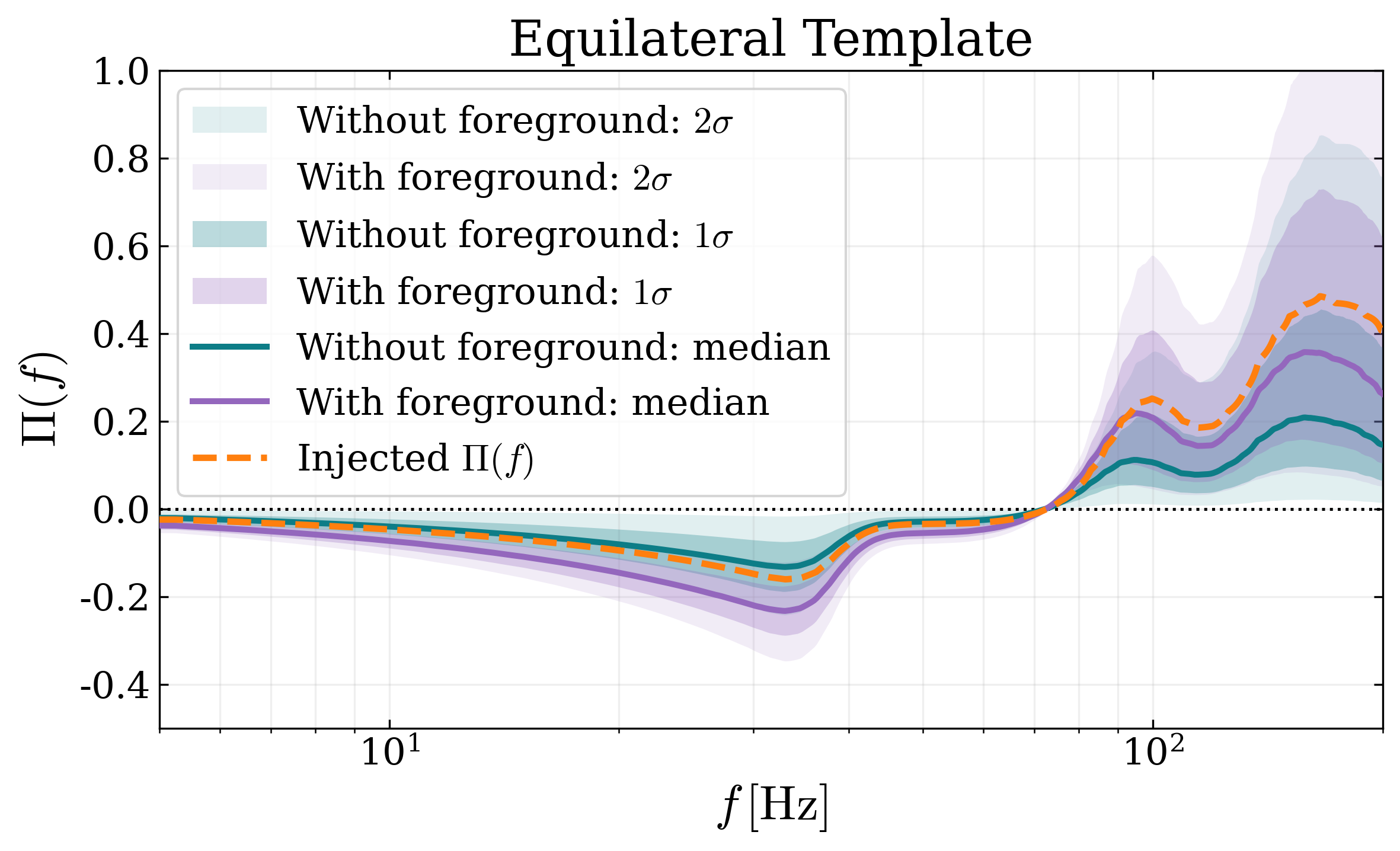}
    \caption{Reconstructed degree of polarization for the equilateral template.} 
    \label{fig:Pi_eq}
\end{figure}

At this point, since we have the posterior estimates for $A_{\rm p}, \, f_{\rm peak}, \, f_{\rm NL}, \, \tau_{\rm NL}, \, \tilde{\tau}_{\rm NL}$, we can build the confidence intervals for $\Pi(f)$.
We present them in Figure \ref{fig:Pi_loc} and \ref{fig:Pi_eq} for the local and equilateral templates, respectively. 
We plot the polarization degree across the frequency range in solid lines,  with the corresponding credible intervals shown as shaded regions. 
We show both the case in which the SIGW signal is analyzed together with the astrophysical foreground, shown in purple, and the case in which only the SIGW contribution is included, shown in teal. 

This method of reconstructing $\Pi(f)$ offers a reliable and physically motivated approach to 
quantify chirality and its scale-dependence, and it generalizes the constant and power-law templates employed for
$\Pi(f)$~\cite{Martinovic:2021hzy,Duval:2025vfg}.

The main message from Figs.~\ref{fig:Pi_loc} and \ref{fig:Pi_eq} is that the reconstructed value of $\Pi(f)$ captures both the small de
gree of chirality in the IR regime, up to the peak of the SIGW spectrum, and the larger chirality in the UV tail.
For the local template, the injected $\Pi(f)$ is recovered within the $1\sigma$ credible band over the relevant frequency range. For the equilateral template, the recovery is less precise: the injected curve remains broadly compatible with the reconstructed posterior bands, but it is inside the $1\sigma$ region only over part of the spectrum and requires a wider $2\sigma$ band in some frequency intervals. This behaviour reflects the broader and more degenerate marginalized posteriors of the parameters entering $\Pi(f)$, as discussed in the previous subsection.

In both the local and equilateral analyses, despite the high uncertainty portrayed in the high-frequency part of the spectrum (namely, larger bounds on the $1\sigma$ region), 
the degree of chirality up to $f=100\,{\rm Hz}$ is well constrained. 
The reason is that below $100\,{\rm Hz}$, the intensity of the spectrum highly dominates over the chiral component (see Figure \ref{fig:ogw-loc}), which as showed in Section~\ref{subsec:PE}, presents smaller relative errors with respect to the chiral one.
 If this on one side shows that we have less precision when having a higher degree of chirality, it also shows that SGWB sources that are theoretically predicted to have mild or small chirality can be constrained using this pipeline with next-generation ground-based detectors. 
 
\section{Discussion}
\label{sec:Discussion}
In addition to the stringent constraints on $A_{\rm p}$ and $f_{\rm peak}$, the tight bounds on $\fnl$ and $\taunl$ highlight the potential of SIGWs as a powerful probe of primordial non-Gaussianity in regimes where scalar perturbations are enhanced to large amplitudes, and where we have poor information from other cosmological probes.
At the same time, the constraint on $\tilde \tau_{\rm NL}$ shows the efficiency of SIGWs in probing primordial parity-violation, especially in the scalar sector.
We should remark that the SIGWs, that can be observed using interferometers such as CE and ET, 
provide one of the very few, possibly unique, probes of primordial scalar non-Gaussianity and 
associated parity violation on small scales of $k = 10^{16}-10^{17}\,\mpcinv$,
corresponding to comoving wavelengths $\lambda \simeq 10^7-10^6\,{\rm m}$.
Any other observable, on such scales in the present universe, is expected to be dominated by local physics, 
with the memory of its primordial initial conditions largely erased by nonlinear evolution.

In the context of the inflationary paradigm, the insights on $A_{\rm p}$, $\fnl$, $\taunl$ and 
$\tilde \tau_{\rm NL}$ over small scales translate to late stage evolution of the inflaton field.
Setting the pivot scale of $5\times 10^{-2}\,\mpcinv$ to have exited the horizon around $50$ e-folds before the end 
of inflation, the constraints over $10^{16}-10^{17}\,\mpcinv$ inform us about inflationary dynamics 
over $8$ to $10$ e-folds before the end of inflation and the onset of reheating.
Thus the parameter constraints contain information about the shape of the underlying inflaton potential close to its minimum,
ending of inflaton's evolution and couplings of inflaton to other fields.
The amplitude of $\tilde \tau_{\rm NL}$ is a clean and unique quantifier of parity-violating interactions in the scalar sector just prior to the end of inflation.
Complementing the observables over large scales such as CMB and the large-scale structure that trace the shape of inflationary potential over early stage evolution of the field, constraints from SIGWs 
hence lend a second anchor to pin-down the shape of inflationary potential and inflaton interactions 
as the field rolls close to the minimum.

\section{Conclusion}
\label{sec:Conclusion}

In this work, we have shown that SIGWs, that can be probed with upcoming interferometers, can constrain primordial parity violation and scalar non-Gaussianity.
We have performed the analysis with a scenario of enhanced scalar power over small scales while accounting for scalar bispectrum and the associated exchange trispectrum containing parity-even and parity-odd components.

We have been able to recover the fiducial parameters within at least $2\sigma$ through Bayesian analysis framework, 
while significantly improving upon the priors on them. Specifically, for the local template of the bispectrum and 
exchange trispectrum, we constrain 
$\fnl = 6.9^{+3.0}_{-4.0}$,\,
$\taunl = 10.2^{+5.4}_{-5.2}$ and 
$\tilde\tau_{\rm NL} = 12.8^{+14.9}_{-14.8}$ along with 
$A_{\rm p}=(9.91^{-0.21}_{-0.24}) \times 10^{-3}$ and $f_{\rm peak} = 50.06 ^{+0.24}_{-0.24}\rm{Hz}$. 
For the equilateral template of the bispectrum and exchange trispectrum, we constrain 
$\fnl = 12.8^{+6.7}_{-7.2}$,\,
$\taunl = 9.6^{+2.9}_{-2.3}$ and 
$\tilde\tau_{\rm NL} = 9.3^{+4.1}_{-3.9}$ along with 
$A_{\rm p}=(9.71^{+0.34}_{-0.47}) \times 10^{-3}$ and $f_{\rm peak} = 49.71 ^{+0.28}_{-0.31}\rm{Hz}$. 
Our results correspond to statistics of primordial perturbations over comoving scales of 
$\mathcal{O}(10^7)\,{\rm m}$.
Such constraints on primordial cosmology over small scales are novel and potentially unique.
In addition, we have also considered a realistic scenario including an astrophysical foreground, 
obtaining compatible results and showing that, despite a small broadening of the posteriors, 
we are still able to constrain such parameters with good precision at $2\sigma$ level.
The unique shape of $\ogw^I$ and $\ogw^V$ arising from SIGW due to a peaked $\ps(k)$ strongly aids in 
distinguishing from the foreground and constraining the underlying primordial parameters.

Our results demonstrate the potential of upcoming interferometric detectors in hunting physics of
early universe and levying new bounds on primordial statistics that are competitive to the existing constraints on large scales.
The constraint on $\fnl$ and $\taunl$ when compared against the existing bounds from CMB may inform us of potential running over a wide range of scales.
Moreover, the constraint on $\tilde \tau_{\rm NL}$ over small scales, in comparison with the current bound from galaxy distribution, shall be a supplementary probe for detecting parity-odd primordial 
scalar trispectrum.

Further, our analysis holds promise for specific model-based analysis, where our constraints can be translated to insights on specific features of the theory, such as the shape of inflaton potential, presence of additional fields, consistency relations between different correlations and strengths of interactions, especially of parity-violating kind.
Moreover, the value of $A_{\rm p} \simeq 10^{-2}$ along with $\fnl$ and $\taunl$ of ${\cal O}(1-10)$, 
over scales of ${\cal O}(10^7)\,{\rm m}$ shall have interesting implications for 
production of PBHs of masses about ${\cal O}(10^{10})\,{\rm kg}$~\cite{Carr:2016drx,Bartolo:2018evs,Carr:2020xqk}.
However, one must be careful in computation of PBH abundance from scalar perturbations with 
substantial non-Gaussianities and associated 
interpretations (see, for instance~\cite{Young:2013oia,Franciolini:2018vbk,Atal:2019cdz,Cai:2022erk,Ferrante:2022mui}).
We intend to pursue some of these directions in future work.

\acknowledgments
\noindent
HVR and NB acknowledge support by the MUR PRIN2022 Project ``BROWSEPOL: Beyond 
standaRd mOdel With coSmic microwavE background POLarization''-2022EJNZ53 financed 
by the European Union - Next Generation EU. 
HVR and NB acknowledge support by the INFN InDark Initiative. 

HVR acknowledges the use of computing cluster under 
University of Padova Strategic Research Infrastructure Grant 2017:
``CAPRI: Calcolo ad Alte Prestazioni per la Ricerca e l’Innovazione''. 
NB acknowledges financial support from the COSMOS network (www.cosmosnet.it)
through the ASI (Italian Space Agency) Grants 2016-24-H.0, 2016-24-H.1-2018 and 2020-9-HH.0. 
\appendix

\section{Expressions of $I$ and $Q$}
\label{app:IQ}

We provide explicit forms of certain functions used in the analysis.
The factor $Q^\lambda(\vk,\vq)$ in Eq.~\eqref{eq:h-RR} is a dimensionless function that 
arises from the contraction of the polarization tensor (for a given polarization $\lambda$)
defined with respect to the propagation direction of $\hat \vk$, and the wavevector 
of integration $\vq$ in the following fashion
\begin{equation}
Q^\lambda(\vk,\vq) = e^\lambda_{ij}(\vk)\f{q_iq_j}{k^2}\,.
\end{equation}
Here $e^\lambda_{ij}(\vk)$ is the polarization tensor that can be written in
terms of the polarization vectors orthogonal to $\hat \vk$, in any basis,
say $\lambda=[+,\times]$ or $\lambda=[L,R]$~\cite{Domenech:2021ztg}.
For reference, if we orient $\vq_z \parallel \vk$, we may compute
\begin{align}
Q^L(\vk,\vq)=[Q^R(\vk,\vq)\vert]^\ast
=& \f{1}{2}\left(\f{q}{k}\right)^2\sin^2\theta\,e^{2i\phi}\,.
\end{align}

The function $\tilde I(a,b,c,d)$ that first appears in Eq.~\eqref{eq:ph-I-G} is a dimensionless
function arising from the evolution of the inhomogeneous solution of the two mode functions 
$h_\vk^\lambda$ constituting the two-point correlation~\cite{Adshead:2021hnm,Perna:2024ehx}.
It results from averaging over fast oscillations in the solution and extracting the overall 
time dependence as $1/(k^2\eta^2)$ from the products of Greens functions in the integrand.
The explicit expression of the function $\tilde I$ is given by~\cite{Adshead:2021hnm}
\begin{align}
\tilde I\left(u_1,v_1,u_2,v_2\right) &=
I_A(u_1,v_1)I_A(u_2,v_2) \bigg[I_B(u_1,v_1)I_B(u_2,v_2) \nn \\
& + \pi^2 I_C(u_1,v_1)I_C(u_2,v_2) \bigg]\,,
\end{align}
where
\begin{align}
I_A(u,v) &= \f{3(u^2+v^2-3)}{4u^3v^3}\,, \\
I_B(u,v) &= -4uv + (u^2+v^2-3)\ln \left\vert \f{3 - (u+v)^2}{3 - (u-v)^2}\right\vert\,, \\
I_C(u,v) &= (u^2+v^2-3)\Theta(u+v-\sqrt{3})\,.
\end{align}

\begin{widetext}
\section{Bispectrum and exchange trispectrum from a given three-point vertex}\label{app:3pt-4pt}
Here, we briefly review the structure of the scalar bispectrum and the exchange trispectrum 
arising from a given three-point interaction and the resultant relation between the parameters
$\fnl$ and $\taunl$.
Let us introduce the non-Gaussian contribution to $\cR$ in Fourier space 
arising from a general three-point interaction
as $\cR_\vk=\cR_\vk^{\rm G} + \cR^{3\rm nG}_\vk$ where
\begin{align}
\cR^{3\rm nG}_\vk &= -\f{3}{5}\fnl\int\f{\d^3\vq}{(2\pi)^3}
w_3(k,q,\vert\vk-\vq\vert)\cR_{\vq} \cR_{\vk-\vq}\,,
\label{eq:Rk3ng}
\end{align}
The motivation is that $\fnl$ captures the dimensionless strength of the given interaction
while the function $w_3$ captures the non-trivial shape and scale dependence of the interaction~\cite{Acquaviva:2002ud,Lyth:2005fi,Schmidt:2010gw,Agullo:2021oqk,Ragavendra:2021qdu}.\footnote{The product of $\fnl w_3(k_1,k_2,k_3)$ is given as $\fnl(k_1,k_2,k_3)$ in Ref.~\cite{Ragavendra:2021qdu}.}
This leads to the three-point correlation to be
\begin{align}
\Braket{\cR_{\vk_1}\cR_{\vk_2}\cR_{\vk_3}} &= -\f{6}{5}(2\pi)^3
\delta^{(3)}(\vk_1+\vk_2+\vk_3)\,
\fnl w_3(k_1,k_2,k_3)\left[P_\cR(k_1)P_\cR(k_2) +~\text{two permutations}\right]\,.
\end{align}
We may then identify the bispectrum ${\cal B}(k_1,k_2,k_3)$ from the above expression through 
the definition $\Braket{\cR_{\vk_1}\cR_{\vk_2}\cR_{\vk_3}}=(2\pi)^3
\delta^{(3)}(\vk_1+\vk_2+\vk_3)\,{\cal B}(k_1,k_2,k_3)$ as
\begin{align}
{\cal B}(k_1,k_2,k_3) &= -2\fnl w_3(k_1,k_2,k_3)\left[P_\cR(k_1)P_\cR(k_2) +~\text{two permutations}\right]\,.
\end{align}
Clearly, $w_3=1$ leads to the local template of bispectrum.
For the equilateral template, we shall have~\cite{Planck:2019kim}
\begin{align}
w_3(k_1,k_2,k_3) =& -3\bigg\{\bigg[P_\cR(k_1) P_\cR(k_2) + P_\cR(k_2)P_\cR(k_3) + P_\cR(k_1)P_\cR(k_3) 
+ 2 \left[P_\cR(k_1)P_\cR(k_2)P_\cR(k_3)\right]^{2/3}\bigg]\nn \\
&- \bigg[P^{1/3}_\cR(k_1) P^{2/3}_\cR(k_2) P_\cR(k_3) + \text{5 permutations} \bigg]\bigg\}
\left[P_\cR(k_1)P_\cR(k_2) + \text{two permutations}\right]^{-1}\,.
\label{eq:w3-eq}
\end{align}

Further substituting $\cR_\vk^{3\rm nG}$ in the four-point correlation
$\Braket{\cR_{\vk_1}\cR_{\vk_2}\cR_{\vk_3}\cR_{\vk_4}}$ and
extracting contractions that lead to connected terms with $\delta(\vk_1+\vk_2+\vk_3+\vk_4)$,
we get
\begin{align}
\Braket{\cR_{\vk_1}\cR_{\vk_2}\cR_{\vk_3}\cR_{\vk_4}} &= 
(4)\f{9}{25}(2\pi)^3\delta^{(3)}(\vk_1+\vk_2+\vk_3+\vk_4)\fnl^2 \nn \\ 
&\bigg\{P_\cR(k_1)P_\cR(k_2)
\bigg[P_\cR(|\vk_3+\vk_1|)w_3(k_1,k_3,\vert\vk_1+\vk_3\vert)w_3(k_2,k_4,\vert\vk_2+\vk_4\vert)
+ (\vk_1 \leftrightarrow \vk_2) \bigg]\nn \\
& + P_\cR(k_2)P_\cR(k_3)
\bigg[P_\cR(|\vk_1+\vk_2|) w_3(k_1,k_2,\vert\vk_1+\vk_2\vert)w_3(k_3,k_4,\vert\vk_3+\vk_4\vert)
+ (\vk_2 \leftrightarrow \vk_3)\bigg]\nn \\
& + P_\cR(k_3)P_\cR(k_4)
\bigg[P_\cR(|\vk_1+\vk_3|) w_3(k_1,k_3,\vert\vk_1+\vk_3\vert)w_3(k_2,k_4,\vert\vk_2+\vk_4\vert)
+ (\vk_3 \leftrightarrow \vk_4)\bigg]\nn \\
& + P_\cR(k_4)P_\cR(k_1)
\bigg[P_\cR(|\vk_2+\vk_4|) w_3(k_2,k_4,\vert\vk_2+\vk_4\vert)w_3(k_3,k_1,\vert\vk_3+\vk_1\vert)
+ (\vk_4 \leftrightarrow \vk_1)\bigg]\nn \\
& + P_\cR(k_1)P_\cR(k_3)
\bigg[P_\cR(|\vk_2+\vk_1|) w_3(k_2,k_1,\vert\vk_2+\vk_1\vert)w_3(k_3,k_4,\vert\vk_3+\vk_4\vert)
+ (\vk_1 \leftrightarrow \vk_3)\bigg]\nn \\
& + P_\cR(k_2)P_\cR(k_4)
\bigg[P_\cR(|\vk_1+\vk_2|) w_3(k_1,k_2,\vert\vk_1+\vk_2\vert)w_3(k_2,k_3,\vert\vk_2+\vk_3\vert)
+ (\vk_2 \leftrightarrow \vk_4)\bigg]\bigg\}
\end{align}
or in shorthand
\begin{align}
\Braket{\cR_{\vk_1}\cR_{\vk_2}\cR_{\vk_3}\cR_{\vk_4}} = 
\f{36}{25}(2\pi)^3\delta^{(3)}(\vk_1+\vk_2+\vk_3+\vk_4)\fnl^2
& \big[P_\cR(k_1)P_\cR(k_2)P_\cR(|\vk_3+\vk_1|)
w_3(k_1,k_3,\vert\vk_1+\vk_3\vert)w_3(k_2,k_4,\vert\vk_2+\vk_4\vert)\nn \\
& +~11\,\text{permutations}\big]\,.
\end{align}
Recall that $P_\cR(k) = 2\pi^2\ps(k)/k^3$.
We may now define the parameter $\taunl$ through the relation
\begin{align}
\Braket{\cR_{\vk_1}\cR_{\vk_2}\cR_{\vk_3}\cR_{\vk_4}} &= (2\pi)^3\taunl
w_4(\vk_1,\vk_2,\vk_3,\vk_4)\delta^{(3)}(\vk_1+\vk_2+\vk_3+\vk_4) 
\left[P_\cR(k_1)P_\cR(k_2)P_\cR(|\vk_3+\vk_1|) +~11\,\text{permutations}\right]\,, 
\end{align}
and identify the shape function of the trispectrum $w_4$ from the previous expression to be
\begin{align}
w_4(\vk_1,\vk_2,\vk_3,\vk_4) &= \f{\left[P_\cR(k_1)P_\cR(k_2)P_\cR(|\vk_3+\vk_1|)
w_3(k_1,k_3,\vert\vk_1+\vk_3\vert)w_3(k_2,k_4,\vert\vk_2+\vk_4\vert)+~11\,\text{permutations}\right]}
{\left[P_\cR(k_1)P_\cR(k_2)P_\cR(|\vk_3+\vk_1|) +~11\,\text{permutations}\right]}\,.
\label{eq:w4_w3}
\end{align}
The local case of $w_3=1$ reduces $w_4=1$.
In essence, the above procedure gives us the shape of the bispectrum and the associated exchange trispectrum from a given three-point interaction.
We may utilize a given template of bispectrum, say local or equilateral, in the above expression and obtain the corresponding template of the exchange trispectrum.

Further, we also see that $\taunl = \left(\f{36}{25}\right)\fnl^2$.
It is the saturation of Suyama-Yamaguchi inequality, as we have considered $\taunl$ and $\fnl$
sourced by the same interaction.
In the presence of other three-point interactions sourcing the exchange trispectrum,
it becomes the inequality of 
$\taunl \geq \left(\f{36}{25}\right)\fnl^2\,.$
Moreover, the inequality may be relaxed in a parity-violating scenario with a non-vanishing 
$\tilde\tau_{_{\rm NL}}$.
To account for the generality of the different three-point contributions to the exchange trispectrum, we treat the overall normalization $\taunl$ as a free parameter.
We retain the relation between $w_4$ and $w_3$ as in Eq.~\eqref{eq:w4_w3}, while in the most
general case, this can also be relaxed.
To arrive at the shape of the odd-part of the exchange trispectrum, we use the same procedure as above with the normalization being $\tilde\tau_{_{\rm NL}}$.
So the parity-even and parity-odd parts of the trispectrum (barring the Dirac delta functions) are
\begin{align}
\Braket{\cR_{\vk_1}\cR_{\vk_2}\cR_{\vk_3}\cR_{\vk_4}}'_{\rm even} &= 
\taunl\left[P_\cR(k_1)P_\cR(k_2)P_\cR(|\vk_3+\vk_1|)
w_3(k_1,k_3,\vert\vk_1+\vk_3\vert)w_3(k_2,k_4,\vert\vk_2+\vk_4\vert)
+~11\,\text{permutations}\right]\,,\\
\Braket{\cR_{\vk_1}\cR_{\vk_2}\cR_{\vk_3}\cR_{\vk_4}}'_{\rm odd} &= i\,
\tilde\tau_{_{\rm NL}}\left[P_\cR(k_1)P_\cR(k_2)P_\cR(|\vk_3+\vk_1|)
w_3(k_1,k_3,\vert\vk_1+\vk_3\vert)w_3(k_2,k_4,\vert\vk_2+\vk_4\vert)
+~11\,\text{permutations}\right]\,.
\end{align}
\color{black}
We use the local case of $w_3=1$ and the equilateral case of $w_3$ as in Eq.~\eqref{eq:w3-eq} 
in our analyses to obtain the constraints on the associated $\fnl$, $\taunl$ and $\tilde \tau_{_{\rm NL}}$.

\subsection{Contributions to SIGW}

Having the non-Gaussian component of scalar perturbation $\cR^{\rm 3nG}$ informing the generic shape
of bispectrum and associated exchange trispectrum, we turn to the contributions to SIGW
which are uniquely arising from the loop-level contribution to the scalar power due to the cubic order interaction.
We substitute $\cR^{\rm 3nG}$ as given in Eq.~\eqref{eq:Rk3ng} in 
$\Braket{\cR_{\vk_1}\cR_{\vk_2}\cR_{\vk_3}\cR_{\vk_4}}$ and collect the terms that contain 
$\delta^{(3)}(\vk_1+\vk_2)\delta^{(3)}(\vk_3+\vk_4)$ i.e.,
\begin{align}
\Braket{\cR_{\vk_1}\cR_{\vk_2}\cR_{\vk_3}\cR_{\vk_4}} =&
\Braket{\cR_{\vk_1}\cR_{\vk_2}}\Braket{\cR^{\rm 3nG}_{\vk_3}\cR^{\rm 3nG}_{\vk_4}} 
+~5\,\text{permutations}\,.
\end{align}
Written explicitly, they are
\begin{align}
\Braket{\cR_{\vk_1}\cR_{\vk_2}\cR_{\vk_3}\cR_{\vk_4}} = 
(2)\f{9}{25}\delta^{(3)}(\vk_1+\vk_2)\delta^{(3)}(\vk_3+\vk_4)\fnl^2
& \bigg[P_\cR(k_1)
\int \f{\d^3\vq}{(2\pi)^3} P_\cR(q) P_\cR(|\vk_3-\vq|)
w_3^2(k_3,q,\vert\vk_3-\vq\vert)\nn \\
& +~5\,\text{permutations}\bigg]\,.
\end{align}
The corresponding scalar-induced tensor power spectrum is~\cite{Unal:2018yaa,Adshead:2021hnm,Ragavendra:2021qdu,Perna:2024ehx}
\begin{align}
\ph^\lambda(k)\bigg\vert_{\rm hybrid} =& 8\left(\f{9}{25}\right)\left(\f{8}{k^2\eta^2}\right)
\f{k^3}{2\pi^2}\fnl^2\iint\f{\d^3\vq_1\d^3\vq_2}{(2\pi)^6}\vert Q^\lambda(\vk,\vq_1)\vert^2
\tilde I\left(\f{q_1}{k},\f{|\vk-\vq_1|}{k},\f{q_1}{k},\f{|\vk-\vq_1|}{k}\right)\nn \\
&\times w_3^2(q_1,q_2,\vert\vq_1-\vq_2\vert)
P_\cR(|\vk-\vq_1|)P_\cR(q_2)P_\cR(|\vq_1-\vq_2|)\,.
\end{align}

Then we collect the terms that contain $\delta^{(3)}(\vk_1+\vk_2+\vk_3+\vk_4)$.
We consider terms where the substitution of $\cR^{\rm 3nG}$ is done on the adjacent pairs of mode 
functions in the four-point correlation, such as $\Braket{\cR^{\rm 3nG}_{\vq_1}\cR^{\rm 3nG}_{\vk-\vq_1};\cR_{-\vq_2}\cR_{\vq_2-\vk}}$, and the contraction shall be between the first pair and the second pair.
They shall be
\begin{align}
\Braket{\cR^{\rm 3nG}_{\vq_1}\cR^{\rm 3nG}_{\vk-\vq_1};\cR_{-\vq_2}\cR_{\vq_2-\vk}}
& = 4\f{9}{25}\delta^{(3)}(\vk-\vk)\fnl^2\,P_\cR(q_2)P_\cR(|\vq_2-\vk|)\nn \\ 
&\times\big[P_\cR(|\vq_1-\vq_2|)w_3(q_1,q_2,\vert\vq_1-\vq_2\vert)
w_3(\vert \vk-\vq_1\vert,\vert q_2-\vk\vert,\vert\vq_2-\vq_1\vert)\nn \\
& + P_\cR(|\vk-\vq_1-\vq_2|)w_3(q_1,\vert\vq_2-\vk\vert,\vert\vq_2-k+\vq_1\vert)
w_3(\vert\vk-\vq_1\vert,q_2,\vert\vk-\vq_1-\vq_2\vert)\big]
\end{align}
and 
\begin{align}
\Braket{\cR_{\vq_1}\cR_{\vk-\vq_1};\cR^{\rm 3nG}_{-\vq_2}\cR^{\rm 3nG}_{\vq_2-\vk}}
& = 4\f{9}{25}\delta^{(3)}(\vk-\vk)\fnl^2\,P_\cR(q_1)P_\cR(|\vk-\vq_1|)\nn \\ 
&\times\big[P_\cR(|\vq_1-\vq_2|)w_3(q_1,q_2,\vert\vq_1-\vq_2\vert)
w_3(\vert \vk-\vq_1\vert,\vert q_2-\vk\vert,\vert\vq_2-\vq_1\vert)\nn \\
& + P_\cR(|\vk-\vq_1-\vq_2|) w_3(q_1,\vert\vq_2-\vk\vert,\vert\vq_2-k+\vq_1\vert)
w_3(\vert\vk-\vq_1\vert,q_2,\vert\vk-\vq_1-\vq_2\vert)\big]\,.
\end{align}
Summing them we get
\begin{align}
\Braket{\cR^{\rm 3nG}_{\vq_1}\cR^{\rm 3nG}_{\vk-\vq_1};\cR_{-\vq_2}\cR_{\vq_2-\vk}}
+& \Braket{\cR_{\vq_1}\cR_{\vk-\vq_1};\cR^{\rm 3nG}_{-\vq_2}\cR^{\rm 3nG}_{\vq_2-\vk}}
= \f{36}{25}\delta^{(3)}(\vk-\vk)\fnl^2\,[P_\cR(q_1)P_\cR(|\vk-\vq_1|)
+P_\cR(q_2)P_\cR(|\vk-\vq_2|)] \nn \\ 
&\times\big[P_\cR(|\vq_1-\vq_2|)w_3(q_1,q_2,\vert\vq_1-\vq_2\vert)
w_3(\vert \vk-\vq_1\vert,\vert q_2-\vk\vert,\vert\vq_2-\vq_1\vert)\nn \\
& + P_\cR(|\vk-\vq_1-\vq_2|) w_3(q_1,\vert\vq_2-\vk\vert,\vert\vq_2-k+\vq_1\vert)
w_3(\vert\vk-\vq_1\vert,q_2,\vert\vk-\vq_1-\vq_2\vert)\big]\,.
\end{align}
This is the C-type contribution to SIGW power spectrum~\cite{Unal:2018yaa,Adshead:2021hnm,Ragavendra:2021qdu}.
\begin{align}
\ph^\lambda(k)\bigg\vert_{\rm C} =& \left(\f{36}{25}\right)\left(\f{8}{k^2\eta^2}\right)
\f{k^3}{2\pi^2}\fnl^2
\iint\f{\d^3\vq_1\d^3\vq_2}{(2\pi)^6}Q^\lambda(\vk,\vq_1)Q^\lambda(-\vk,-\vq_2)
\tilde I\left(\f{q_1}{k},\f{|\vk-\vq_1|}{k},\f{q_2}{k},\f{|\vk-\vq_2|}{k}\right)\nn \\
& \times [P_\cR(q_1)P_\cR(|\vk-\vq_1|)+P_\cR(q_2)P_\cR(|\vk-\vq_2|)]\nn \\
& \times \bigg[P_\cR(|\vq_1-\vq_2|)\,w_3(q_1,q_2,\vert\vq_1-\vq_2\vert)
w_3(\vert \vk-\vq_1\vert,\vert q_2-\vk\vert,\vert\vq_2-\vq_1\vert)\nn \\
& + P_\cR(\vk-\vq_1-\vq_2)\,w_3(q_1,\vert\vq_2-\vk\vert,\vert\vq_2-k+\vq_1\vert)
w_3(\vert\vk-\vq_1\vert,q_2,\vert\vk-\vq_1-\vq_2\vert)\bigg]\,.
\end{align}
Using the symmetry of the integrand under $\vq_1 \to \vk-\vq_1$, we can show
that the two terms within the final pair of square braces are equivalent.
So
\begin{align}
\ph^\lambda(k)\bigg\vert_{\rm C} = 
2\left(\f{36}{25}\right)\left(\f{8}{k^2\eta^2}\right)\f{k^3}{2\pi^2}\fnl^2
\iint & \f{\d^3\vq_1\d^3\vq_2}{(2\pi)^6}Q^\lambda(\vk,\vq_1)Q^\lambda(-\vk,-\vq_2)
\tilde I\left(\f{q_1}{k},\f{|\vk-\vq_1|}{k},\f{q_2}{k},\f{|\vk-\vq_2|}{k}\right)\nn \\
& \times w_3(q_1,q_2,\vert\vq_1-\vq_2\vert)
w_3(\vert \vk-\vq_1\vert,\vert q_2-\vk\vert,\vert\vq_2-\vq_1\vert)\nn \\
& \times P_\cR(|\vq_1-\vq_2|)\,[P_\cR(q_1)P_\cR(|\vk-\vq_1|) + P_\cR(q_2)P_\cR(|\vk-\vq_2|)]\,.
\end{align}

We then turn to terms where the substitution of $\cR^{\rm 3nG}$ is on the non-adjacent mode 
functions in the four-point correlation as 
$\Braket{\cR^{\rm 3nG}_{\vq_1}\cR_{\vk-\vq_1};\cR^{\rm 3nG}_{-\vq_2}\cR_{\vq_2-\vk}}$
and the contraction is between the first pair and the second pair of mode functions.
They shall be
\begin{align}
\Braket{\cR^{\rm 3nG}_{\vq_1}\cR_{\vk-\vq_1};\cR^{\rm 3nG}_{-\vq_2}\cR_{\vq_2-\vk}}
=& \f{36}{25}\delta^{(3)}(\vk-\vk)\fnl^2 P_\cR(|\vk-\vq_1|)P_\cR(|\vq_2-\vk|)
\bigg[P_\cR(k)\,w_3(q_1,\vert\vk-\vq_1\vert,k)w_3(q_2,\vert q_2-\vk\vert,k)\nn \\
& + P_\cR(|\vk-\vq_1-\vq_2|)\,w_3(q_1,\vert\vq_2-\vk\vert,\vert\vq_2+\vq_1-\vk\vert)
w_3(\vert\vk-\vq_1\vert,q_2,\vert\vk-\vq_1-\vq_2\vert)\bigg]\,,\\
\Braket{\cR_{\vq_1}\cR_{\vk-\vq_1}^{\rm 3nG};\cR^{\rm 3nG}_{-\vq_2}\cR_{\vq_2-\vk}}
=& \f{36}{25}\delta^{(3)}(\vk-\vk)\fnl^2 P_\cR(q_1)P_\cR(|\vq_2-\vk|)
\bigg[P_\cR(k)\,w_3(q_1,\vert\vk-\vq_1\vert,k)w_3(q_2,\vert q_2-\vk\vert,k)\nn \\
& + P_\cR(|\vq_2-\vq_1|)\,w_3(\vert\vk-\vq_1\vert,\vert\vq_2-\vk\vert,\vert\vq_2-\vq_1\vert)
w_3(q_1,q_2,\vert\vq_2-\vq_1\vert)\bigg]\,,\\
\Braket{\cR_{\vq_1}\cR_{\vk-\vq_1}^{\rm 3nG};\cR_{-\vq_2}\cR^{\rm 3nG}_{\vq_2-\vk}}
=& \f{36}{25}\delta^{(3)}(\vk-\vk)\fnl^2 P_\cR(q_1)P_\cR(q_2)
\bigg[P_\cR(k)\,w_3(q_1,\vert\vk-\vq_1\vert,k)w_3(q_2,\vert q_2-\vk\vert,k)\nn \\
& + P_\cR(|\vk-\vq_1-\vq_2|)w_3(q_1,\vert\vq_2-\vk\vert,\vert\vq_2+\vq_1-\vk\vert)
w_3(\vert\vk-\vq_1\vert,q_2,\vert\vk-\vq_1-\vq_2\vert)\bigg]\,,\\
\Braket{\cR^{\rm 3nG}_{\vq_1}\cR_{\vk-\vq_1};\cR_{-\vq_2}\cR^{\rm 3nG}_{\vq_2-\vk}}
=& \f{36}{25}\delta^{(3)}(\vk-\vk)\fnl^2 P_\cR(|\vk-\vq_1|) P_\cR(q_2)
\bigg[P_\cR(k)w_3(q_1,\vert\vk-\vq_1\vert,k)w_3(q_2,\vert q_2-\vk\vert,k)\nn \\
& + P_\cR(|\vq_2-\vq_1|)\,w_3(q_1,q_2,\vert\vq_2-\vq_1\vert)
w_3(\vert\vk-\vq_1\vert,\vert\vq_2-\vk\vert,\vert\vq_2-\vq_1\vert)\bigg]\,.
\end{align}
Summing the four terms above, we have
\begin{align}
\begin{rcases}
\Braket{\cR^{\rm 3nG}_{\vq_1}\cR_{\vk-\vq_1};\cR^{\rm 3nG}_{-\vq_2}\cR_{\vq_2-\vk}}\nn\\
+\Braket{\cR^{\rm 3nG}_{\vq_1}\cR_{\vk-\vq_1};\cR_{-\vq_2}\cR^{\rm 3nG}_{\vq_2-\vk}}\nn \\
+\Braket{\cR_{\vq_1}\cR^{\rm 3nG}_{\vk-\vq_1};\cR^{\rm 3nG}_{-\vq_2}\cR_{\vq_2-\vk}}\nn\\
+\Braket{\cR_{\vq_1}\cR^{\rm 3nG}_{\vk-\vq_1};\cR_{-\vq_2}\cR^{\rm 3nG}_{\vq_2-\vk}}
\end{rcases}
=& \f{36}{25}\delta^{(3)}(\vk-\vk)P_\cR(k)\fnl^2w_3(q_1,\vert\vk-\vq_1\vert,k)
w_3(q_2,\vert q_2-\vk\vert,k)\nn\\
&\times \big[P_\cR(q_1)P_\cR(q_2) + P_\cR(q_1)P_\cR(|\vk-\vq_2|) \nn \\
& + P_\cR(|\vk-\vq_1|) P_\cR(q_2) + P_\cR(|\vk-\vq_1|) P_\cR(|\vk-\vq_2|) \big] \nn \\
+ & \f{36}{25}\delta^{(3)}(\vk-\vk) \fnl^2\bigg[w_3(q_1,\vert\vq_2-\vk\vert,\vert \vq_1+\vq_2-\vk\vert) w_3(\vert \vk-\vq_1\vert,q_2,\vert \vk -\vq_1-\vq_2\vert)\nn \\
& \times P_\cR(|\vk-\vq_1-\vq_2|) 
\big[P_\cR(q_1)P_\cR(q_2) + P_\cR(|\vk-\vq_1|) P_\cR(|\vk-\vq_2|)\big] \nn \\
& + w_3(\vert \vk-\vq_1\vert,\vert \vq_2-\vk\vert,\vert \vq_2-\vq_1\vert)
w_3(q_1,\vq_2,\vert\vq_1-\vq_2\vert) \nn\\
& \times P_\cR(|\vq_1-\vq_2|)
\big[P_\cR(q_1)P_\cR(|\vq_2-\vk|) + P_\cR(|\vk-\vq_1|)P_\cR(q_2)\big]\bigg]\,.
\end{align}
The first set of terms (in the first three lines of RHS) proportional to $P_\cR(k)$ vanishes 
upon integration over $\vq_1,\,\vq_2$ (azimuthal integration over $\phi_2-\phi_1$).
This is the C-type contribution to SIGW power spectrum~\cite{Unal:2018yaa,Adshead:2021hnm,Ragavendra:2021qdu}.
\begin{align}
\ph^\lambda(k)\bigg\vert_{\rm Z} =& \left(\f{36}{25}\right)\left(\f{8}{k^2\eta^2}\right)
\f{k^3}{2\pi^2}\fnl^2
\iint\f{\d^3\vq_1\d^3\vq_2}{(2\pi)^6}Q^\lambda(\vk,\vq_1)Q^\lambda(-\vk,-\vq_2)
\tilde I\left(\f{q_1}{k},\f{|\vk-\vq_1|}{k},\f{q_2}{k},\f{|\vk-\vq_2|}{k}\right)\nn \\
& \times \bigg[w_3(q_1,q_2,\vert\vq_1-\vq_2\vert)
w_3(\vert \vk-\vq_1\vert,\vert q_2-\vk\vert,\vert\vq_2-\vq_1\vert)\nn \\
& \times P_\cR(|\vq_1-\vq_2|)
[P_\cR(q_1)P_\cR(|\vq_2-\vk|) + P_\cR(|\vk-\vq_1|) P_\cR(q_2)]\nn \\
& + w_3(q_1,\vert\vq_2-\vk\vert,\vert\vq_2-k+\vq_1\vert)
w_3(\vert\vk-\vq_1\vert,q_2,\vert\vk-\vq_1-\vq_2\vert)\nn \\
& \times P_\cR(|\vk-\vq_1-\vq_2|)
[P_\cR(q_1)P_\cR(q_2) + P_\cR(|\vk-\vq_1|)P_\cR(|\vq_2-\vk|)]\bigg]\,.
\end{align}

Just to consolidate, summing all six type of substitutions above leading to C-type and Z-type
contributions, and using the symmetry of the integrand under $\vq_1 \leftrightarrow \vk-\vq_1$,
we may simplify them as
\begin{align}
\begin{rcases}
\Braket{\cR^{\rm 3nG}_{\vq_1}\cR^{\rm 3nG}_{\vk-\vq_1};\cR_{-\vq_2}\cR_{\vq_2-\vk}}
+\Braket{\cR_{\vq_1}\cR_{\vk-\vq_1};\cR^{\rm 3nG}_{-\vq_2}\cR^{\rm 3nG}_{\vq_2-\vk}}\nn \\
+\Braket{\cR^{\rm 3nG}_{\vq_1}\cR_{\vk-\vq_1};\cR^{\rm 3nG}_{-\vq_2}\cR_{\vq_2-\vk}}
+\Braket{\cR^{\rm 3nG}_{\vq_1}\cR_{\vk-\vq_1};\cR_{-\vq_2}\cR^{\rm 3nG}_{\vq_2-\vk}}\nn \\
+\Braket{\cR_{\vq_1}\cR^{\rm 3nG}_{\vk-\vq_1};\cR^{\rm 3nG}_{-\vq_2}\cR_{\vq_2-\vk}}
+\Braket{\cR_{\vq_1}\cR^{\rm 3nG}_{\vk-\vq_1};\cR_{-\vq_2}\cR^{\rm 3nG}_{\vq_2-\vk}}
\end{rcases}
= & 2\left(\f{36}{25}\right)\delta^{(3)}(\vk-\vk)
\fnl^2 w_3(q_1,q_2,\vert\vq_1-\vq_2\vert)\nn \\
& \times w_3(\vert \vk-\vq_1\vert,\vert q_2-\vk\vert,\vert\vq_2-\vq_1\vert) 
P_\cR(|\vq_1-\vq_2|)\nn \\
& \times \big[P_\cR(q_1)P_\cR(|\vk-\vq_1|) + P_\cR(\vq_2) P_\cR(|\vk-\vq_2|) \nn \\
& + P_\cR(q_1) P_\cR(|\vq_2-\vk|) + P_\cR(|\vk-\vq_1|) P_\cR(q_2)\big]\,.
\end{align}
Further, using symmetry of the integrand under $\vq_1 \leftrightarrow \vq_2$
\begin{align}
\begin{rcases}
\Braket{\cR^{\rm 3nG}_{\vq_1}\cR^{\rm 3nG}_{\vk-\vq_1};\cR_{-\vq_2}\cR_{\vq_2-\vk}}
+\Braket{\cR_{\vq_1}\cR_{\vk-\vq_1};\cR^{\rm 3nG}_{-\vq_2}\cR^{\rm 3nG}_{\vq_2-\vk}}\nn \\
+\Braket{\cR^{\rm 3nG}_{\vq_1}\cR_{\vk-\vq_1};\cR^{\rm 3nG}_{-\vq_2}\cR_{\vq_2-\vk}}
+\Braket{\cR^{\rm 3nG}_{\vq_1}\cR_{\vk-\vq_1};\cR_{-\vq_2}\cR^{\rm 3nG}_{\vq_2-\vk}}\nn \\
+\Braket{\cR_{\vq_1}\cR^{\rm 3nG}_{\vk-\vq_1};\cR^{\rm 3nG}_{-\vq_2}\cR_{\vq_2-\vk}}
+\Braket{\cR_{\vq_1}\cR^{\rm 3nG}_{\vk-\vq_1};\cR_{-\vq_2}\cR^{\rm 3nG}_{\vq_2-\vk}}
\end{rcases}
= & 4\left(\f{36}{25}\right)\delta^{(3)}(\vk-\vk)\fnl^2 
w_3(q_1,q_2,\vert\vq_1-\vq_2\vert)\nn \\
& \times w_3(\vert \vk-\vq_1\vert,\vert q_2-\vk\vert,\vert\vq_2-\vq_1\vert)
P_\cR(|\vq_1-\vq_2|) \nn \\
& \times \big[P_\cR(q_1)P_\cR(|\vk-\vq_1|) + P_\cR(q_1) P_\cR(|\vq_2-\vk|)\big]\,.
\end{align}
Thus the corresponding scalar-induced tensor power per $\lambda$ can be written compactly
as
\begin{align}
\ph^\lambda(k)\bigg\vert_{\rm C+Z} =& 4\left(\f{36}{25}\right)\left(\f{8}{k^2\eta^2}\right)
\f{k^3}{2\pi^2}\fnl^2
\iint\f{\d^3\vq_1\d^3\vq_2}{(2\pi)^6}Q^\lambda(\vk,\vq_1)Q^\lambda(-\vk,-\vq_2)
\tilde I\left(\f{q_1}{k},\f{|\vk-\vq_1|}{k},\f{q_2}{k},\f{|\vk-\vq_2|}{k}\right)\nn \\
&\times w_3(q_1,q_2,\vert\vq_1-\vq_2\vert)
w_3(\vert \vk-\vq_1\vert,\vert q_2-\vk\vert,\vert\vq_2-\vq_1\vert)\nn \\
& \times P_\cR(q_1)P_\cR(|\vq_1-\vq_2|)
[P_\cR(|\vk-\vq_1|) + P_\cR(|\vq_2-\vk|)]\,.
\end{align}
Generalizing to the contribution from an exchange trispectrum of multiple three-point vertices
of a given shape, we may replace the overall normalization of $(36/25)\fnl^2$ with $\taunl$
and obtain
\begin{align}
\ph^\lambda(k)\bigg\vert_{\rm C+Z} =& 4\left(\f{8}{k^2\eta^2}\right)
\f{k^3}{2\pi^2}\taunl
\iint\f{\d^3\vq_1\d^3\vq_2}{(2\pi)^6}Q^\lambda(\vk,\vq_1)Q^\lambda(-\vk,-\vq_2)
\tilde I\left(\f{q_1}{k},\f{|\vk-\vq_1|}{k},\f{q_2}{k},\f{|\vk-\vq_2|}{k}\right)\nn \\
&\times w_3(q_1,q_2,\vert\vq_1-\vq_2\vert)
w_3(\vert \vk-\vq_1\vert,\vert q_2-\vk\vert,\vert\vq_2-\vq_1\vert)\nn \\
& \times P_\cR(q_1)P_\cR(|\vq_1-\vq_2|)
[P_\cR(|\vk-\vq_1|) + P_\cR(|\vq_2-\vk|)]\,.
\end{align}
Hence we have the expression of $\ph^\lambda(k)$ arising from the exchange 
trispectrum, respecting the symmetries of integration.
The combinatorial factor is $4$, two from C-type and two from Z-type diagrams.

Collecting the relevant expressions
\begin{align}
\ph^{\lambda}(k)\bigg\vert_{\rm Gaussian} &= 2\left(\f{8}{k^2\eta^2}\right)\f{k^3}{2\pi^2}
\int \f{\d^3\vq}{(2\pi)^3} \vert Q^\lambda(\vk,\vq)\vert^2
\tilde I\left(\f{q}{k},\f{|\vk-\vq|}{k},\f{q}{k},\f{|\vk-\vq|}{k}\right)
P_\cR(q) P_\cR(|\vk-\vq|)\,,\\
\ph^\lambda(k)\bigg\vert_{\rm hybrid} =& 8\left(\f{9}{25}\right)\left(\f{8}{k^2\eta^2}\right)
\f{k^3}{2\pi^2}\fnl^2\iint\f{\d^3\vq_1\d^3\vq_2}{(2\pi)^6}\vert Q^\lambda(\vk,\vq_1)\vert^2
\tilde I\left(\f{q_1}{k},\f{|\vk-\vq_1|}{k},\f{q_1}{k},\f{|\vk-\vq_1|}{k}\right)\nn \\
&\times w_3^2(q_1,q_2,\vert\vq_1-\vq_2\vert)
P_\cR(|\vk-\vq_1|)P_\cR(q_2)P_\cR(|\vq_1-\vq_2|)\,,\\
\ph^\lambda(k)\bigg\vert_{\rm C+Z} =& 4\left(\f{8}{k^2\eta^2}\right)\f{k^3}{2\pi^2}\taunl
\iint\f{\d^3\vq_1\d^3\vq_2}{(2\pi)^6}Q^\lambda(\vk,\vq_1)Q^\lambda(-\vk,-\vq_2)
\tilde I\left(\f{q_1}{k},\f{|\vk-\vq_1|}{k},\f{q_2}{k},\f{|\vk-\vq_2|}{k}\right)\nn \\
&\times w_3(q_1,q_2,\vert\vq_1-\vq_2\vert)
w_3(\vert \vk-\vq_1\vert,\vert q_2-\vk\vert,\vert\vq_2-\vq_1\vert)
P_\cR(q_1)P_\cR(|\vq_1-\vq_2|)
[P_\cR(|\vk-\vq_1|) + P_\cR(|\vq_2-\vk|)]\,.
\end{align}
Once again, the C-type and Z-type terms arise from exchange trispectrum and so are connected contributions,
whereas the hybrid term arises from partially disconnected contribution.

The power spectrum of intensity is $\ph^I(k) = \ph^L(k)+\ph^R(k)$.
We may note that
$\vert Q^L(\vk,\vq)\vert^2=\vert Q^R(\vk,\vq)\vert^2=(1/4)(q/k)^4\sin^4\theta$
[cf.~App.~\ref{app:IQ}].
So, we may obtain $\ph^I$ from each of the contributions listed above as
\begin{align}
\ph^I(k)\bigg\vert_{\rm Gaussian} =& 2(2)\left(\f{8}{k^2\eta^2}\right)\f{k^3}{2\pi^2}
\int \f{\d^3\vq}{(2\pi)^3} \vert Q^L(\vk,\vq)\vert^2
\tilde I\left(\f{q}{k},\f{|\vk-\vq|}{k},\f{q}{k},\f{|\vk-\vq|}{k}\right)
P_\cR(q) P_\cR(|\vk-\vq|) \\
\ph^I(k)\bigg\vert_{\rm hybrid} =& 2(8)\left(\f{9}{25}\right)\left(\f{8}{k^2\eta^2}\right)
\f{k^3}{2\pi^2}\fnl^2 \iint\f{\d^3\vq_1\d^3\vq_2}{(2\pi)^6}\vert Q^L(\vk,\vq_1)\vert^2
\tilde I\left(\f{q_1}{k},\f{|\vk-\vq_1|}{k},\f{q_1}{k},\f{|\vk-\vq_1|}{k}\right)\nn \\
&\times w_3^2(q_1,q_2,\vert\vq_1-\vq_2\vert)
P_\cR(|\vk-\vq_1|)P_\cR(q_2)P_\cR(|\vq_1-\vq_2|)\,,\\
\ph^I(k)\bigg\vert_{\rm C + Z} =& 4\left(\f{8}{k^2\eta^2}\right)
\f{k^3}{2\pi^2}\taunl \iint\f{\d^3\vq_1\d^3\vq_2}{(2\pi)^6}
[Q^L(\vk,\vq_1)Q^L(-\vk,-\vq_2)
+ Q^R(\vk,\vq_1)Q^R(-\vk,-\vq_2)]\nn \\
& \times \tilde I\left(\f{q_1}{k},\f{|\vk-\vq_1|}{k},\f{q_2}{k},\f{|\vk-\vq_2|}{k}\right)
w_3(q_1,q_2,\vert\vq_1-\vq_2\vert) 
w_3(\vert \vk-\vq_1\vert,\vert q_2-\vk\vert,\vert\vq_2-\vq_1\vert) \nn \\
&\times P_\cR(q_1)P_\cR(|\vq_1-\vq_2|)
[P_\cR(|\vk-\vq_1|) + P_\cR(|\vq_2-\vk|)]\,.
\end{align}

The power spectrum of V mode is
$\ph^V(k) = \ph^L(k)-\ph^R(k)$.
So $\ph^V(k)\big\vert_{\rm Gaussian} = 
\ph^V(k)\big\vert_{\rm hybrid} = 0$ and
\begin{align}
\ph^V(k)\bigg\vert_{\rm C + Z} =& 4\,i\,\left(\f{8}{k^2\eta^2}\right)
\f{k^3}{2\pi^2} \tilde\tau_{_{\rm NL}} \iint\f{\d^3\vq_1\d^3\vq_2}{(2\pi)^6}
[Q^L(\vk,\vq_1)Q^L(-\vk,-\vq_2) - Q^R(\vk,\vq_1)Q^R(-\vk,-\vq_2)]\nn \\
& \times \tilde I\left(\f{q_1}{k},\f{|\vk-\vq_1|}{k},\f{q_2}{k},\f{|\vk-\vq_2|}{k}\right)
w_3(q_1,q_2,\vert\vq_1-\vq_2\vert) 
w_3(\vert \vk-\vq_1\vert,\vert q_2-\vk\vert,\vert\vq_2-\vq_1\vert) \nn \\
&\times P_\cR(q_1)P_\cR(|\vq_1-\vq_2|)
[P_\cR(|\vk-\vq_1|) + P_\cR(|\vq_2-\vk|)]\,.
\end{align}
Note that only $C$ and $Z$ type terms contribute to V-mode as they are the leading order
contributions sensitive to trispectrum.
Further, the combination of polarization factors $[Q^L(\vk,\vq_1)Q^L(-\vk,-\vq_2) - Q^R(\vk,\vq_1)Q^R(-\vk,-\vq_2)]$ are fully imaginary and it extracts out the parity-odd part of the trispectrum,
which is imaginary too (hence the $i$ in the prefactor). 
The resulting $\ph^V(k)$ shall be real and captures parity-violation.
\end{widetext}

\bibliographystyle{apsrev4-2}
\bibliography{bibliography}

\end{document}